\documentclass{aastex62}

\usepackage{natbib}

\usepackage{graphicx}

\usepackage{amssymb}

\usepackage{geometry}
\usepackage{color}
\usepackage{wasysym}

\graphicspath{{FIGURES/}}

\begin{document}




\title{Interior properties of the inner Saturnian moons from space astrometry data\footnote{10.1016/j.icarus.2019.01.026} \footnote {© 2019. This manuscript version is made available under the CC-BY-NC-ND 4.0 license http://creativecommons.org/licenses/by-nc-nd/4.0/}}








\author{V. Lainey} 
\affiliation{Jet Propulsion Laboratory, California Institute of Technology, 4800 Oak Grove Drive, Pasadena, CA 91109-8099, United States}
\affiliation{IMCCE, Observatoire de Paris, PSL Research University, CNRS,  Sorbonne Universit\'e, Univ. Lille, 77 Avenue Denfert-Rochereau, 75014 Paris, France}
\author{B. Noyelles}
\affiliation{NAmur Institute for CompleX SYStems, University of Namur, Rempart de la Vierge 8, B-5000 Namur, Belgium}
\affiliation{Institut UTINAM, CNRS UMR 6213, Univ. Bourgogne Franche-Comt\'e, OSU THETA, BP 1615, F-25010 Besan\c{c}on Cedex, France}
\author{N. Cooper}
\affiliation{Queen Mary University of London, Mile End Rd, London E1 4NS, United Kingdom}
\affiliation{IMCCE, Observatoire de Paris, PSL Research University, CNRS,  Sorbonne Universit\'e, Univ. Lille, 77 Avenue Denfert-Rochereau, 75014 Paris, France}
\author{N. Rambaux}
\affiliation{IMCCE, Observatoire de Paris, PSL Research University, CNRS,  Sorbonne Universit\'e, Univ. Lille, 77 Avenue Denfert-Rochereau, 75014 Paris, France}
\author{C. Murray}
\affiliation{Queen Mary University of London, Mile End Rd, London E1 4NS, United Kingdom}
\author{R. S. Park}
\affiliation{Jet Propulsion Laboratory, California Institute of Technology, 4800 Oak Grove Drive, Pasadena, CA 91109-8099, United States}

\correspondingauthor{V.Lainey}
\email{valery.j.lainey@jpl.nasa.gov}


\begin{abstract}
During the thirteen years in orbit around Saturn before its final plunge, the Cassini spacecraft provided more than ten thousand astrometric measurements. Such large amounts of accurate data enable the search for extremely faint signals in the orbital motion of the saturnian moons. Among these, the detection of the dynamical feedback of the rotation of the inner moons of Saturn on their respective orbits becomes possible. Using all the currently available astrometric data associated with Atlas, Prometheus, Pandora, Janus and Epimetheus, we first provide a detailed analysis of the Cassini Imaging Science Subsystem (ISS) data, with special emphasis on their statistical behavior and sources of bias. Then, we give updated estimates of the moons' averaged densities and try to infer more details about their interior properties by estimating the physical librations for Prometheus, Pandora, Epimetheus and Janus from anomalies in their apsidal precession. Our results are compatible with a homogeneous interior for Janus and Epimetheus, within the uncertainty of the measurements. On the other hand, we found some inconsistency for Pandora and Prometheus, which might result from a dynamical mismodeling of Saturn's gravity field. Last, we show how the synergistic introduction of libration measurements directly derived from imaging should allow the moons' moments of inertia to be better constrained.
\end{abstract}

\keywords{Inner Saturnian moons, interior, astrometry}



\section{Introduction}

Assuming an age of less than tens of millions of years \citep{Char10}, the inner moons of Saturn are witnesses of the recent evolution of the Saturnian system. As a consequence of the exchange of angular momentum between the moons and Saturn's rings, the system was more compact in the past \citep{1982AJ.....87.1051L}. This suggested that the inner moons may have formed near the outer edge of the Roche limit by accretion of icy ring material \citep{Char10}. Such a scenario is supported by the monotonic distribution of the moons' masses as a function of distance to their primary, as well as their small densities of 0.4-0.7 g/cm$^3$ \citep{Thom13}. Since collisions between moonlets are expected during such a formation process (\citet{Char10}, \citet{2018NatAs...2..555L}), a possible inhomogeneity may arise within the interior of these small icy moons. Here, we try to assess the interior distribution of these small moons by determining their physical librations. 
Using Cassini ISS data, \citet{Tisc09} studied the physical librations of Janus and Epimetheus by measuring the change in rotation of the two moons as a function of time. They quantified the physical libration of Epimetheus to $-5.9\pm 1.2^{\circ}$, but could not reach a conclusive result for Janus. Here, we consider a different approach that relies on the dynamical feedback of the physical libration into the orbit. In particular, while the measurement of faint librations requires imaging at close range, precise orbital monitoring relies more on the time interval covered by the data, since we are looking for a secular effect. Benefitting from thousands of ISS astrometric measurements, the fit of a physical libration parameter simultaneously with initial state vectors of the moons and other physical parameters may provide an effective way of characterizing physical librations of small objects. In the next section we provide a short overview of physical librations. Section \ref{sec:feedback} gives details about the quantification of physical librations from orbital motion and the dynamical model we used. Section \ref{sec:obs} presents the observations. Section \ref{sec:fit} presents the results after fitting our orbital model to the astrometric data, as well as a solution in which the quantification of \citet{Tisc09} is introduced as an extra observational constraint. The last section discusses the results.

\section{Interior and physical librations}

In common with most of the natural satellites of the giant planets, the inner moons are assumed to rotate synchronously, as was confirmed for Janus and Epimetheus by Tiscareno et al. (2009). The variations of the Saturn-satellite distance arising due to their orbital eccentricity, and the departure from a spherical shape, induce a varying gravitational torque and associated longitudinal librations as for the main satellites \citep[e.g.][]{Com03, Ram14}.
\newline

There are at least two ways to define the longitudinal librations: i) the \emph{tidal librations} $\psi$, which represent the East-West oscillations of the long axis of the satellite with respect to the planet-satellite direction; 
ii) the \emph{physical librations} $\gamma$, representing the departure between the actual rotation and one with a constant rate $\omega$ (which would be equal to the orbital frequency $n$).
These two quantities differ by the \emph{optical libration} $2e\sin (nt)$, where $e$ is the orbital eccentricity. If the optical libration has just a kinematic origin, the physical and tidal librations contain information on the interior of the rotating object. In particular, we have (e.g. \citet{Murr99}, Eq.~5.123):

\begin{equation}
\label{eq:physical}
\gamma(t) = \frac{2e}{1-(n/\omega_u)^2}\sin(nt)+O(e^2),
\end{equation}
and (Tiscareno et al. 2009, Eq.~15):

\begin{equation}
\label{eq:tidal}
\psi(t) = \frac{-2e}{1-(\omega_u/n)^2}\sin(nt)+O(e^2),
\end{equation}
$\omega_u$ being the frequency of the proper oscillations of the longitudinal motion. This frequency contains a signature of the interior from

\begin{equation}
\label{eq:omegau}
\omega_u = n\sqrt{3\frac{B-A}{C}} = 2n\sqrt{\frac{3C_{22}}{C/(mR^2)}},
\end{equation}
where $m$ is the mass of the satellite, $R$ its mean radius, and $A$, $B$ and $C$ the 3 principal moments of inertia (associated respectively with body-centered
axes $\hat{x}$, $\hat{y}$ and $\hat{z}$). Assuming a homogeneous density $\rho$, we have

\begin{eqnarray}
A & = & \iiint_V \rho\,(y^2+z^2)\,\textrm{d}x\,\textrm{d}y\,\textrm{d}z, \label{eq:GA} \\
B & = & \iiint_V \rho\,(x^2+z^2)\,\textrm{d}x\,\textrm{d}y\,\textrm{d}z, \label{eq:GB} \\
C & = & \iiint_V \rho\,(x^2+y^2)\,\textrm{d}x\,\textrm{d}y\,\textrm{d}z, \label{eq:GC}
\end{eqnarray}
and $C_{22}$ is the classical Stokes coefficient defined as

\begin{equation}
\label{eq:C22}
C_{22} = \frac{B-A}{4mR^2}.
\end{equation}

\par In usual cases, where $(B-A)/C \ll 1/3$, $\omega_u$ is very small with respect to $n$, which results in $\gamma \propto e\frac{B-A}{C}$, while the 
tidal libration $\psi$ is widely dominated by the optical libration for mid and large size moons. This is why it is usually advisable to consider the physical libration $\gamma$ as a direct signature of the triaxiality of the satellite. In the following, we denote its amplitude as $\mathcal{A}$, i.e.

\begin{equation}
\label{eq:physical2}
\gamma(t) = \mathcal{A}\sin(nt)+O(e^2).
\end{equation}
\newline

If the satellite is a homogeneous triaxial ellipsoid with radii $a$, $b$ and $c$, we have

\begin{eqnarray}
A & = & \frac{m}{5}\left(b^2+c^2\right), \label{eq:GA2} \\
B & = & \frac{m}{5}\left(a^2+c^2\right), \label{eq:GB2} \\
C & = & \frac{m}{5}\left(a^2+b^2\right), \label{eq:GC2}
\end{eqnarray}
and

\begin{equation}
\label{eq:ampli}
\mathcal{A} = 3e\frac{a^2-b^2}{a^2-2b^2}.
\end{equation}
\newline

The moments of inertia are computed assuming homogeneous interiors and using the best-fit ellipsoids from \citet{Thom13}, which are only approximations to the shapes for Atlas, Prometheus, Pandora, and Janus. In cases where a digitized shape model is available, we have used it, as in the case of Epimetheus from \citet{Tisc09}.
Table ~\ref{tab:shapes} summarizes the shapes and expected amplitudes of librations of the inner moons Atlas, Prometheus, Pandora, Epimetheus, and Janus. 
The Table also provides the geometric (also called epicyclic) eccentricities, and not the Keplerian ones (\citet{Gree81} and \citet{2006CeMDA..94..237R}). These eccentricites 
and the other orbital elements are derived in considering the oblateness of Saturn. We have in particular 

\begin{equation}
	\label{es:kepler2}
	n^2a^3 = GM_{\saturn}\left(1+\frac{3}{2}J_2\left(\frac{R_{\saturn}}{a}\right)^2+O(J_4)\right).
\end{equation}

The geometric elements have the advantage of not being polluted by the short-period oscillations, which result from the interaction with the flattening of the central body.
We note that negative amplitudes are expected, except for Prometheus. As explained in \citet{Tisc09}, a positive amplitude is a consequence of the very elongated shape of a satellite.
\newline

\begin{table}
\centering
\caption{Best-fit ellipsoid shapes, geometrical orbital eccentricities, and theoretical physical librations of
	some inner moons of Saturn. The shapes are taken from \citet{Thom13}, the eccentricites are the mean epicyclic ones, computed from our model, and the librations are calculated
	using Eq.~\ref{eq:ampli}, except for Epimetheus for which \citet{Tisc09} have shown a significant discrepancy between this method and the actual libration. We used the ratio of 
moments of inertia $(B-A)/C=0.296^{+0.019}_{-0.027}$ instead, derived by Ibid. for the observed shape. The $1 \sigma$ uncertainties were calculated assuming no correlations between the parameters.\label{tab:shapes}}

\begin{tabular}{l|ccccccc}
Satellite & $a$ (km) & $b$ (km) & $c$ (km) & $R$ (km) & $e$ & $\mathcal{A} (^{\circ})$ \\
\hline
Atlas      &  $20.5\pm0.9$ & $17.8\pm0.7$ &  $9.4\pm0.8$ & $15.1\pm0.8$ & $1.1\times10^{-3}$ & $-0.09\pm0.06$ \\
Prometheus &  $68.2\pm0.8$ & $41.6\pm1.8$ & $28.2\pm0.8$ & $43.1\pm1.2$ & $2.2\times10^{-3}$ & $0.93\pm0.19$ \\
Pandora    &  $52.2\pm1.8$ & $40.8\pm2.0$ & $31.5\pm0.9$ & $40.6\pm1.5$ & $4.2\times10^{-3}$ & $-1.27\pm1.07$ \\
Epimetheus &  $(64.9\pm1.3)$ & $(57.3\pm2.5)$ & $(53.0\pm0.5)$ & $(58.2\pm1.2)$ & $9.69\times10^{-3}$ & $-8.9^{-10.4}_{+4.2}$ \\
Janus      & $101.7\pm1.6$ & $93.0\pm0.7$ & $76.3\pm0.4$ & $89.2\pm0.8$ & $6.77\times10^{-3}$ & $-0.29\pm0.08$ \\
\hline
\end{tabular}
\end{table}

For the inner Saturnian moons, the only detected amplitude so far reported in the literature, is $(-5.9\pm1.2)^{\circ}$ for Epimetheus \citep{Tisc09}.
In this specific case, the analysis can be pushed a bit further. In particular, Epimetheus is in 1:1 orbital resonance with Janus implying a horseshoe-shaped orbit. This peculiar orbital motion involves a switch of a few tens of kilometers in the semi-major axes of the two satellites that perturbs the librational motion. As a consequence, their mean motion cannot be considered to be constant. In order to describe the effect of this switch on the librational motion and to extract all the information contained in the libration amplitude, \citet{Robu11} have developed a quasi-periodic librational model where the librations $\gamma$ are decomposed as:

\begin{eqnarray}
	\gamma & =  & \sum_{1\leq  q \leq N}   \beta_q\sin\left( q\nu t + \varphi_q\right)+\frac{ 2e_g \omega_u^2J_0(\beta_1)}{\omega_u^2 - n_g^2} \sin{( n_g t + \ell_0)}  \nonumber \\
	& + &   2e_g \omega_u^2\sum_{p \geq 1}   J_p(\beta_1) \left[\frac{\sin( (n_g +p\nu )t +  p\varphi_1 + \ell_0)}{\omega_u^2 - (n_g +p\nu )^2} + (-1)^p\frac{\sin( (n_g -p\nu)t - p\varphi_1 + \ell_0)}{\omega_u^2 - (n_g - p\nu )^2}\right].
\label{eq:sol_forced}
\end{eqnarray}

The first sum represents the long-period term directly related to the switching motion. In this sum $\beta_q$ are the coefficients of the quasi-periodic series from the variations of the mean-longitude, $\nu$ the frequency of the swap, and $\varphi_q$ its phase. 
The short period librations are described by the second and third terms. Their frequencies are of the order of $n_g$, the geometrical mean motion, with harmonics in $\nu$. The constant $e_g$ is the geometrical eccentricity, and $J_p$ are the Bessel functions.

\section{Measuring the moon's rotation feedback onto the orbit}\label{sec:feedback}

\subsection{Orbital characterization of physical libration}

Benefiting from detailed images of Phobos obtained with the Viking 1 and 2 spacecraft, \citet{Bord90} looked at the physical consequences of the extended gravity field of Phobos on its rotation and orbit. Among other things, they noticed very interestingly that the effect of Phobos' physical libration on its orbit would provide a secular drift on its periapsis distance that should barely be absorbable within other unknown quantities (e.g. initial state, Mars' quadrupole...) during orbit fitting. Indeed, the node and periapsis drifts are often nearly equal in magnitude, but opposite in sign, making any absorption of the extra secular drift in the gravity field of Mars problematic due to an unavoidable ambiguity on the node.
\citet{Bord90} estimated Phobos' periapsis drift $ae\dot{\varpi}$ to be roughly 500 m/yr, which was significant even at the time, given the accuracy of the astrometric observations. They claimed on p.247: ``This term could provide a more accurate estimate of the libration than direct observation of the figure oscillation from ranging to a lander if all effects acting on apse and node are accounted for and if the lander survives longer than about a year.'' Unfortunately, they could not extract the signal from the astrometric residuals.
\newline

Surprisingly, the effect of Phobos' libration on its orbit did not draw much attention at that time, despite the publication of Phobos 2 astrometric data \citep{Koly91}. Fifteen years later, using Mars Express data, \citet{Lain06} observed an anomaly in the astrometric residuals of Phobos. They determined that such an anomaly could be absorbed into a non-physical (i.e negative $C_{22}$) value of the Phobos gravity field, but did not consider physical libration in their model. Finally, \citet{Jaco10} isolated Phobos's physical libration as the dynamical source of the problem and provided the first estimation of Phobos's physical libration from its orbital motion.
\newline

The secular drift, associated with the quadrupole field of Phobos on its periapsis (or any moon in somewhat similar configuration) was given by Borderies and Yoder (1990) and recalled by Jacobson (2010) to be (see also App.~\ref{app:derivation})
\begin{eqnarray}
\Delta \varpi=\frac{3}{2}\left(\frac{R}{a}\right)^2\left[J_2-2C_{22}\left(5-\frac{4\cal{A}}{e}\right)\right]nt+\frac{3}{2}\left(\frac{R}{a}\right)^2\frac{(J_2+6C_{22})}{e}\sin(M)\label{eq:libr}
\end{eqnarray}
where $a, e, n, M$ and $\varpi$ are the traditional osculating Keplerian semi-major axis, eccentricity, mean motion and mean anomaly, $R, J_2$ and $C_{22}$ denote the mean radius and un-normalized gravity coefficients of Phobos. We see here explicitly the action of the libration amplitude $\mathcal{A}$ on the periapsis. Our $\mathcal{A}$ is $-\mathcal{A}$ in \citet{Jaco10}. Of course, this action of the quadrupole field of the satellite is just one effect affecting the pericenter. The precession of the pericenter is mainly due to the the flattening of Saturn, and to the interactions with the other satellites.

In the following, we used the same approach as \citet{Jaco10}, but applied to Atlas, Prometheus, Pandora, Janus and Epimetheus. It is noteworthy that homogeneity requires information on the three moments of inertia $A, B, C$, while orbital monitoring provides essentially just one piece of information $\Delta \varpi$. We will discuss this in more detail in Section \ref{sec:Disc}.
\newline

The coefficients $J_2$ and $C_{22}$ we use are derived from

\begin{eqnarray}
	J_2 & = & \frac{2c^2-a^2-b^2}{10R^2} \label{eq:c20} \\
	C_{22} & = & \frac{a^2-b^2}{20R^2}, \label{eq:c22b}
\end{eqnarray}
these formulae being themselves derived from

\begin{equation}
	\label{eq:J2}
	J_2 = \frac{A+B-2C}{2mR^2}
\end{equation}
and Eq.~\ref{eq:C22}. Their numerical values are given in Table ~\ref{tab:shapes2}.

\begin{table}
	\centering
	\caption{Stokes coefficients $J_2$ and $C_{22}$ for the inner moons. These numbers are derived from the radii given in Tab.~\ref{tab:shapes} in the Eqs.~\ref{eq:c20} and \ref{eq:c22b}.\label{tab:shapes2}
}
\begin{tabular}{l|cc}
	Satellite & $J_2$  & $C_{22}$ \\
\hline
	Atlas      &  $0.245765536599272$ & $2.267663698960571\times10^{-2}$ \\
	Prometheus &  $0.257929274713207$ & $7.861391788373231\times10^{-2}$ \\
	Pandora    &  $0.145901380766337$ & $3.215923705986556\times10^{-2}$ \\
	Epimetheus &  $5.542270403042008\times10^{-2}$ & $1.370909649153886\times10^{-2}$ \\
	Janus      & $9.235694966719619\times10^{-2}$ & $1.064450572100787\times10^{-2}$ \\
\hline
\end{tabular}
\end{table}

\subsection{Dynamical modeling}\label{sec:dyn}

Our approach benefited from the NOE numerical code that was successfully applied to the Mars, Jupiter, and Uranus systems \citep{Lain07,Lain08,Lain09}. It is a gravitational N-body code that incorporates highly sensitive modeling and can generate the partial derivatives needed to fit initial positions, velocities, and many global parameters (i.e. masses, $C_{np}$ and $S_{np}$ gravity coefficients, polar orientation, etc.) to the data. For these simulations, the code included (i) gravitational interactions up to degree two in the spherical harmonic expansion of the gravitational potential for the satellites (assuming homogeneous interior, see Table \ref{tab:shapes2}) and up to degree 6 for Saturn; (ii) the perturbations from the eight main moons of Saturn (Mimas, Enceladus, Tethys, Dione Rhea, Titan, Hyperion and Iapetus) using ephemerides from \citet{Lain17}; (iii) the perturbations of the Sun (including the inner planets and the Moon by incorporating their mass in the Solar one) and Jupiter using DE430 ephemerides; (iv) Saturn's precession; (v) Saturn's nutation using SPICE (Acton, 1996; https://naif.jpl.nasa.gov/naif/) kernel header values SAT382.bsp; (vi) the tidal effects introduced by means of the Love number $k_2$ (dissipation was neglected) from \citet{Lain17}; (vii) relativistic corrections. The dynamical equations of motion were numerically integrated in a Saturn-centric frame with inertial axes (conveniently the Earth mean equator of J2000). 
\newline

The equation of motion for a satellite $P_i$ can be expressed as (Lainey et al. 2002, 2007)

\begin{eqnarray}
{\bf \ddot{r}}_i&=&-G(m_0+m_i)\left(\frac{{\bf r}_i}{r_i^3}-{\bf \nabla}_iU_{\bar{\imath}\hat{0}}+{\bf \nabla}_0U_{\bar{0}\hat{\imath}}\right) \nonumber \\
	& + & \sum_{j=1,j\neq i}^{\cal N}Gm_j\left(\frac{{\bf r}_j-{\bf r}_i}{r_{ij}^3}-\frac{{\bf r}_j}{r_j^3}+{\bf \nabla}_jU_{\bar{\jmath}\hat{0}}-{\bf \nabla}_0U_{\bar{0}\hat{\jmath}}-{\bf \nabla}_jU_{\bar{\jmath}\hat{\imath}}+{\bf \nabla}_iU_{\bar{\imath}\hat{\jmath}}\right)\nonumber \\
&+&\frac{(m_0+m_i)}{m_im_0}{\bf F}_{\bar {\imath}\hat 0}^T+\sum_{j=1,j\ne i}^N\left(-\frac{{\bf F}_{\bar {\jmath}\hat 0}^T}{m_0}+\frac{{\bf F}^T_{ij}}{m_i}\right)+GR
\end{eqnarray}
Here,  ${\bf r}_i$ and ${\bf r}_j$ are the position vectors of the satellite $P_i$ and a body $P_j$ (another satellite, the Sun, or Jupiter) with mass $m_j$, subscript $0$ denotes Saturn, $U_{\bar k\hat l}$ is the oblateness gravity field of body $P_l$ at the position of body $P_k$, $GR$ are corrections due to General Relativity \citep{Newh83}, ${\bf F}^T_{\bar l \hat 0}$ the force received by $P_l$ from the tides it raises on its primary and ${\bf F}^T_{ij}$ the mirroring effects of tides raised by one moon on Saturn acting on another moon. 
Neglecting tidal dissipation, these last two forces are equal to \citep{Lain07,Lain17}  

\begin{eqnarray}
{\bf F}^T_{\bar l \hat 0}&=&-\frac{3k_2Gm^2(E_r)^5{\bf r}_{l}}{r_{l}^8},\\
{\bf F}^T_{ij}&=&\frac{3k_2Gm_jm_iR^5}{2r^5_i r^5_j}\left[-\frac{5({\bf r}_i\cdot {\bf r}_j)^2{\bf r}_i}{r^2_i}+r^2_j{\bf r}_i+2({\bf r}_i\cdot {\bf r}_j){\bf r}_j\right].
\end{eqnarray}
\newline

The physical libration $\gamma$ of the moon $P_i$ arises implicitly in the expression of ${\bf \nabla}_0U_{\bar 0\hat{\imath}}$ and ${\bf \nabla}_jU_{\bar {\jmath}\hat{\imath}}$. While these latter expressions can be found in several references \citep[e.g.]{Pete81,Lain02}, the way physical libration can be introduced in such equations is generally not addressed. Hence, we recall that from the elliptic expansion of true anomaly \citep{Murr99,Jaco10}, and denoting $E$ and $v$ for the eccentric and true anomaly respectively, we get
\begin{eqnarray}
\psi&=&-2e\sin(M)+{\cal A}\sin(M)+O(e^2)=-{\cal B}e\sin(M)+O(e^2)\label{eq:librationNOE}\\
v-M&=&2e\sin(M)+\frac{5}{4}e^2\sin(2M)+O(e^3)=2e\sin(E)+\frac{1}{4}e^2\sin(2E)+O(e^3)
\end{eqnarray}

Hence, one can approximate the introduction of the sine of mean anomaly by the sine of the eccentric anomaly. In this last case, we can easily go back to cartesian coordinates with the expression
\begin{eqnarray}
e\sin(E)=\frac{\bf{r}\cdot\bf{v}}{|\bf{r}\times\bf{v}|}+O(e^2)
\end{eqnarray}
For practical reasons, our code introduces ${\cal B}$ as an adjustable quantity instead of ${\cal A}$. One can easily come back to this latter quantity by introducing the averaged eccentricity value. It is noteworthy to mention that despite the sometimes large differences between geometric and keplerian eccentricities for some of the inner moons, the quantity $e\sin(E)$ remains practically unchanged whether computed with geometric or keplerian elements.
\newline

In addition, we checked the accuracy of such a representation by comparing this approach for Epimetheus with the more complete model given in Eq. \ref{eq:sol_forced}. To do this, we performed a least-squares fit of the short-period analytical model (Eq.~\ref{eq:sol_forced}) to the first order model, corresponding to $\mathcal{A} e(t) \sin{M(t)}$ used in our astrometric fit (Eq.~\ref{eq:libr}).  Our comparison showed an agreement between both representations up to 0.15 degrees. Such a modelling accuracy is certainly sufficient considering the present precision of the measurements on Epimetheus' $\mathcal{A}$ value (see Table \ref{tab:sigmaComp}).
\newline

For an unspecified parameter $c_l$ of the model that shall be fitted (e.g. ${\bf r}(t_0), d{\bf r}/dt(t_0), Q$...), a useful relation is (\citet{Lain12} and references therein) 
\begin{eqnarray} 
\frac{\partial}{\partial c_l}\left(\frac{d^2{\bf r}_i}{dt^2}\right)&=&\frac{1}{m_i}\left[\sum_j\left(\frac{\partial {\bf F}_i}{\partial{\bf
r}_j}\frac{\partial{\bf r}_j}{\partial c_l}+\frac{\partial{\bf F}_i}{\partial\dot{{\bf r}}_j}\frac{\partial\dot{{\bf r}}_j}{\partial c_l}
\right)+\frac{\partial{\bf F}_i}{\partial c_l}\right], \label{eq:EQV}
\end{eqnarray} 
where ${\bf F}_i$ is the right hand side of Eq. (1) multiplied by $m_i$. Partial derivatives of the solutions with respect to initial positions and velocities of the satellites and dynamical parameters were computed from simultaneous integration of Eqs. (15) and (21) .
\newline

All our simulations involved solving simultaneously (at least) for the initial state vectors and masses of the moons, the masses and gravity field of Saturn including zonal harmonics $J_2, J_4, J_6$, the orientation and precession of Saturn. Some attempts were made to solve for more parameters (see further sections). No constraints were introduced in the fit, except most of the time for $J_6$ (see further sections) which was constrained at the value estimated by \citet{Jaco06} assuming a 1-$\sigma$ uncertainty of $10^{-4}$ (that is an order of magnitude larger than the published uncertainty).

\section{Observations}\label{sec:obs}

\subsection{Cassini ISS data}

A campaign of astrometric observations of the small inner satellites of Saturn using the Cassini Imaging Science Subsystem (ISS) 
was carried throughout the Cassini tour, between 2004 and 2017 \citep{Porc04,Coop15}. The observations fitted in this work 
come mainly from these so-called SATELLORB image sequences, with the addition of some opportunistic detections of the satellites 
of interest (Atlas, Prometheus, Pandora, Janus and Epimetheus) in other Cassini ISS image sequences designed to study Saturn's F ring. The total
number of Cassini observations used in this work is 3476, where the large majority were observed using the narrow angle camera (NAC) of the ISS.
\newline

For the Cassini ISS data, astrometric data reduction was performed using the Caviar software package \citep[e.g.]{Coop18}.
The nominal camera pointing direction for each image was corrected using background reference stars from the UCAC2 
\citep{Zach04} and Tycho-2 \citep{Hog00} catalogues. We estimate a typical camera pointing accuracy 
of $\simeq$ 0.1 pixel (0.12 arcsec) (see \citet{Coop15} for further discussion). The pixel coordinates corresponding 
to the astrometric position of each satellite in a given image were measured using either the centroid \citet{Stet87} for unresolved images, 
or a limb-fitting method for resolved satellite images. Centroids were corrected from the measured centre-of light to the centre-of-figure using 
a correction for solar phase angle (observer, object, Sun). The limb-fitting technique involved fitting an ellipsoidal model of the satellite limb, 
projected onto the image, to the detected limb, based on an edge-detection method (see \citet{Coop18} for more details).  Ellipsoidal 
shape models were extracted from the SPICE kernels current at the time of reduction \citep[Tab.~1]{Coop15}.
\newline

ISS data have a known bias in the Solar direction associated with the estimation of the limb of the satellites on the image and their 
respective ellipse fitting \citep{Taje13,Coop14}. Unfortunately, no real solution has been provided so far. Moreover, the statistical 
properties of the residuals and especially their Gaussian behavior has not been studied in much detail. Also, the previous studies considered the main 
moons of Saturn, which are close to spherical, while the inner moons considered in this work generally have much more complex shapes. 
In this paper we show that, once a bias in the Solar direction is removed (see \ref{app:1}), the NAC ISS residuals exhibit a behavior that is close to Gaussian (see \ref{app:2}). Importantly, the solar direction bias does not appear to be related to a center of mass / center of figure shift.
\newline

Measurement uncertainties are discussed at length in several previous publications on Cassini astrometry \citep[e.g.]{Taje13}.  For example, the typical uncertainty in the spacecraft position is $\sim$ 100 m, which is two orders of magnitude smaller than the uncertainty in the measured satellite positions, based on limb fitting and given the typical image resolution ($\sim$ 5-10 km/pixel). Also, the image sequences used in this work were designed specifically for astrometry and dynamical modelling, with exposure and filter settings chosen to optimize imaging of background stars. As a result, the typical pointing uncertainty is 0.1 pixel, an order of magnitude smaller than the uncertainty in the measured satellite centers, based on limb-fitting or centroiding. In common with other previous work, our approach has therefore been to absorb these much smaller effects into the overall uncertainty of the measured satellite positions.
\newline

ISS data are provided as sample and line coordinates, with the origin referenced to an edge of the images. Since different cameras and different methods for finding centers of figure were used for the images, we separated the data set into subsets. We noticed that observations performed prior to June 2004 were somewhat biased due to the uncertainty in the spacecraft position. Since only the centroiding technique was used in this case, we put those observations in another separated subset (in the latter sections of the text we refer to these observations as pre-arrival observations). The pre-arrival observations are not numerous, representing only 238 observations out of 3476. The large distance at which these observations were performed make them not very useful from a dynamical constraint point of view, but we kept them for completeness. 

\subsection{Hubble Space Telescope data}

While the ISS data is the essential data set we used in this study, we also considered the Hubble Space Telescope (HST) data as extra data for checking the robustness of our solutions. 
These data were provided by \citet{Fren03,Fren06} in the form of astrometric observations between the end of 1994 and the beginning of 2005.
While this dataset does complement the ISS data, it does not provide a strong constraint on the physical libration, as will be shown in Section \ref{sec:fit}. Moreover, it does not include astrometric data for Atlas, which was too faint to be measured.

\section{Fitting procedure and results}\label{sec:fit}

\subsection{Weighting the data}

Each ISS dataset comes with an uncertainty in pixel units in the center of figure determination. This comes as a by-product of the Caviar software. Nevertheless, uncertainty in the camera pointing is not considered, and more importantly, the real meaning of the Caviar uncertainty as a standard deviation (i.e. $1 \sigma$) is not granted. As a consequence, such information cannot be used as is. In Cooper et al. (2015), all data were weighted assuming a 1 pixel uncertainty. Here we tried to estimate as accurately as possible a $1 \sigma$ uncertainty for each data point. We started by considering as independent random variables, the error on each specific satellite and coordinate (sample or line) data. After performing an initial fit of our dynamical model, we computed the number of observations beyond $3 \sigma$ level. Despite the large number of data points (typically several hundred per satellite, see also Table \ref{tab:moysig}), we found only one observation above $3 \sigma$. As was confirmed later on, this indicated that the uncertainties given in the ISS data are pessimistic. Hence, to assess a better weight, we rescaled the Caviar uncertainties by a scalar chosen to get about 66\% of astrometric residuals within $1 \sigma$ level for each satellite and coordinate. 
\newline

Another issue that arose from the use of ISS data is the redundancy of information coming from different observation series. Indeed, it frequently happened that more than one ISS data point existed within a small time span (less than 10 minutes). This is a real problem when weighting the data since biases in the data series are frequent. In particular, the Solar bias may not be the only bias when considering a single series. For instance, a bias coming from the repetition of a specific geometry (recall that the inner moons have strongly distorted shapes) may prevent a proper convergence of the whole fitting procedure. To prevent this issue we de-weigthed the series data by a factor $\sqrt{N}$ with $N$ being the number of observations in the series. In the full dataset, we found 14, 17, 14, 2 and 3 series respectively for Atlas, Prometheus, Pandora, Epimetheus and Janus.
\newline

ISS data were then weighted using the root-mean-square (rms) of the astrometric residuals, coming from a first estimate of the moons' positions with our model, as an \textit{a priori} constraint when fitting the data. The chaotic orbital motion of these moons required extra care when iterating using a least-squares method. We found it useful to stabilize the convergence process by updating the weights every other iteration only. A few of such double-step iterations were sufficient to achieve a stable estimation.
To prevent chaos spreading, no fits were done using inter-satellite measurements. After the first converged solution, observations with residuals larger than 2.5 pixels were discarded. Only then, observations with residuals higher than three-sigma were removed.
\newline

While ISS data are provided as sample and line, HST data are provided as tangential coordinates on the celestial sphere, the origin of the frame being Saturn's center, after fitting a ring profile during the image calibration. Since different cameras were used for the images, we separated the data set into subsets so that each one was associated with
a different camera. Similarly to the ISS case, data have been weighted using the root-mean-square (rms) of the astrometric residuals as an \textit{a priori} constraint when fitting the data. Here again, no fits were done using inter-satellite measurements. Three-sigma observations were removed.

\subsection{Postfit residuals and fitted parameters}

In the present work, we performed many different fits, each time changing the number of fitted parameters or observation set to check the reliability of the uncertainty of our measurements. 
Most simulations considered ISS data only, since adding HST data was CPU intensive and did not bring much improvement. Over the ten different simulations we used, seven of them were performed using slightly different modeling.
While most of our solutions considered NOE ephemerides for the main moons as given by Lainey et al. (2017), we considered two solutions with JPL's ephemerides sat359l and sat375. Three other simulations tested a different model by adding odd harmonics of Saturn's gravity, or by neglecting the nutations in the fit. Another solution added Pallene's observations, increasing the number of studied moons from five to six. This was found useful, since the inner moons are influenced by Saturn's rings. Adding Pallene, which is evolving further away from the rings and has very little mass, helped to verify that the results were not significantly affected by the absence of the rings in the model. All in all, the different solutions provided similar results within the error bars and suggested that our estimation of fitted parameters is extremely reliable.
\newline

\begin{figure*} 
\begin{center} 
\begin{tabular}{ll} 
\hspace*{-0.35cm}\includegraphics[width=7.5cm,angle=0]{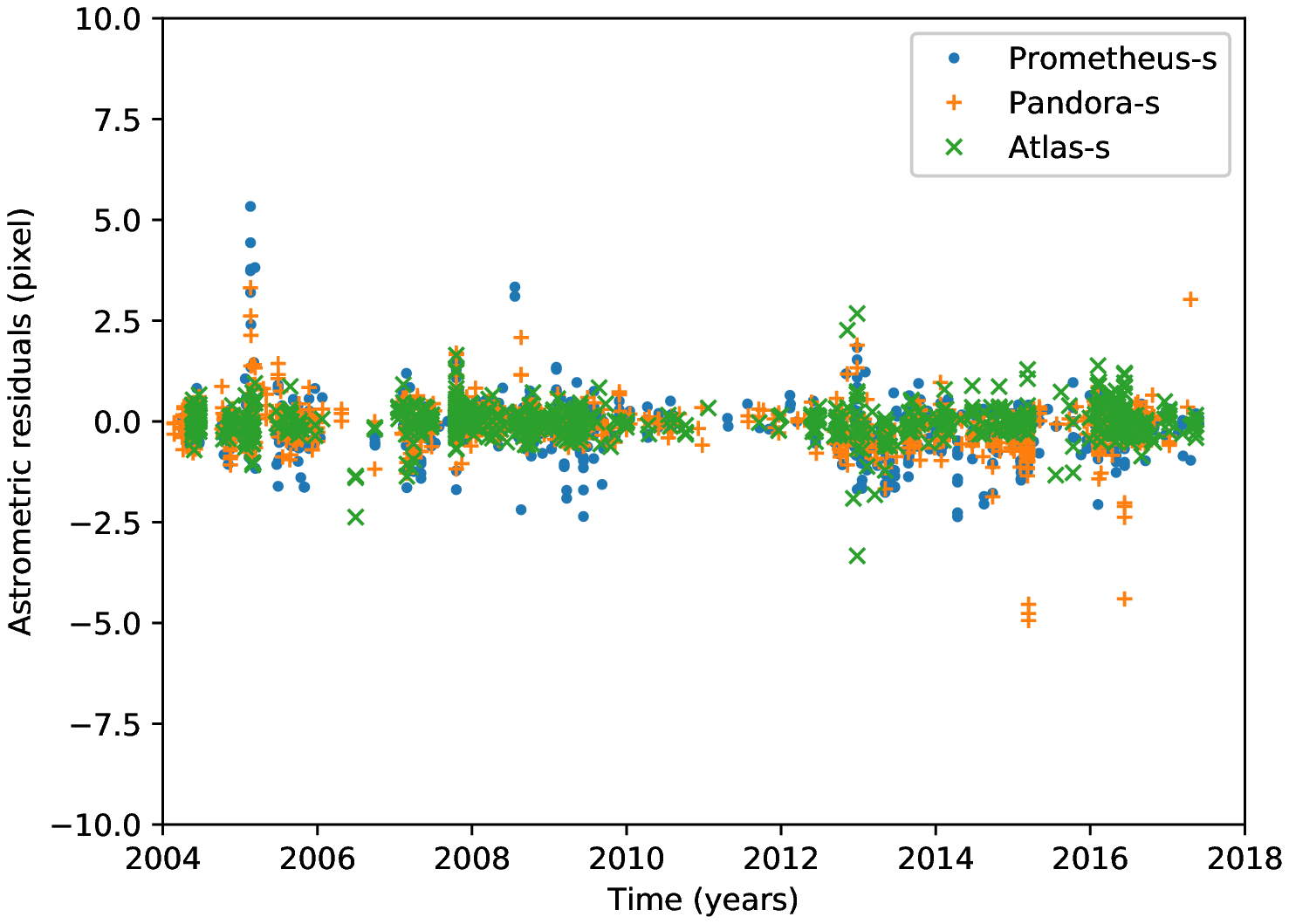} & \includegraphics[width=7.5cm,angle=0]{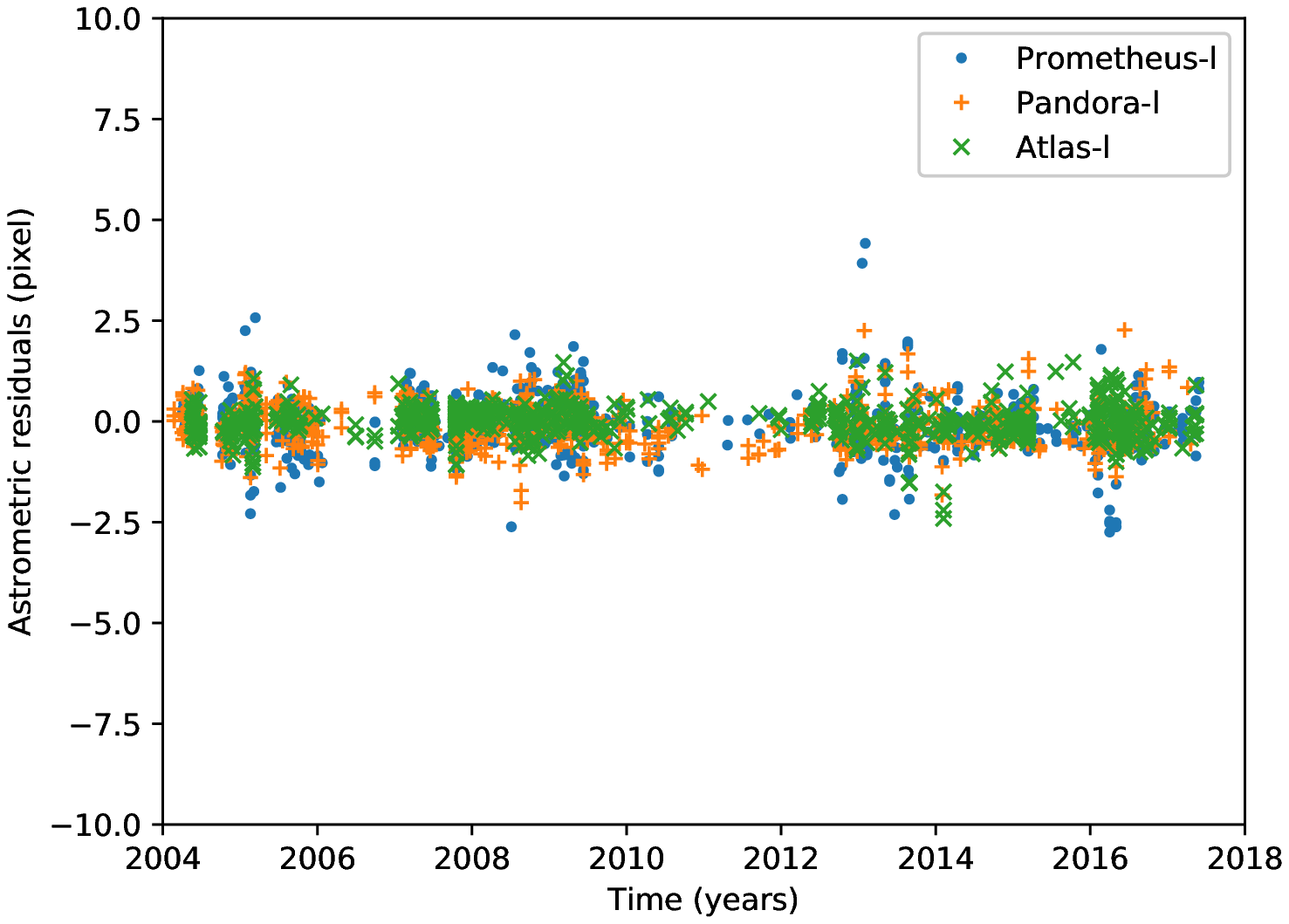}\\
\hspace*{-0.35cm}\includegraphics[width=7.5cm,angle=0]{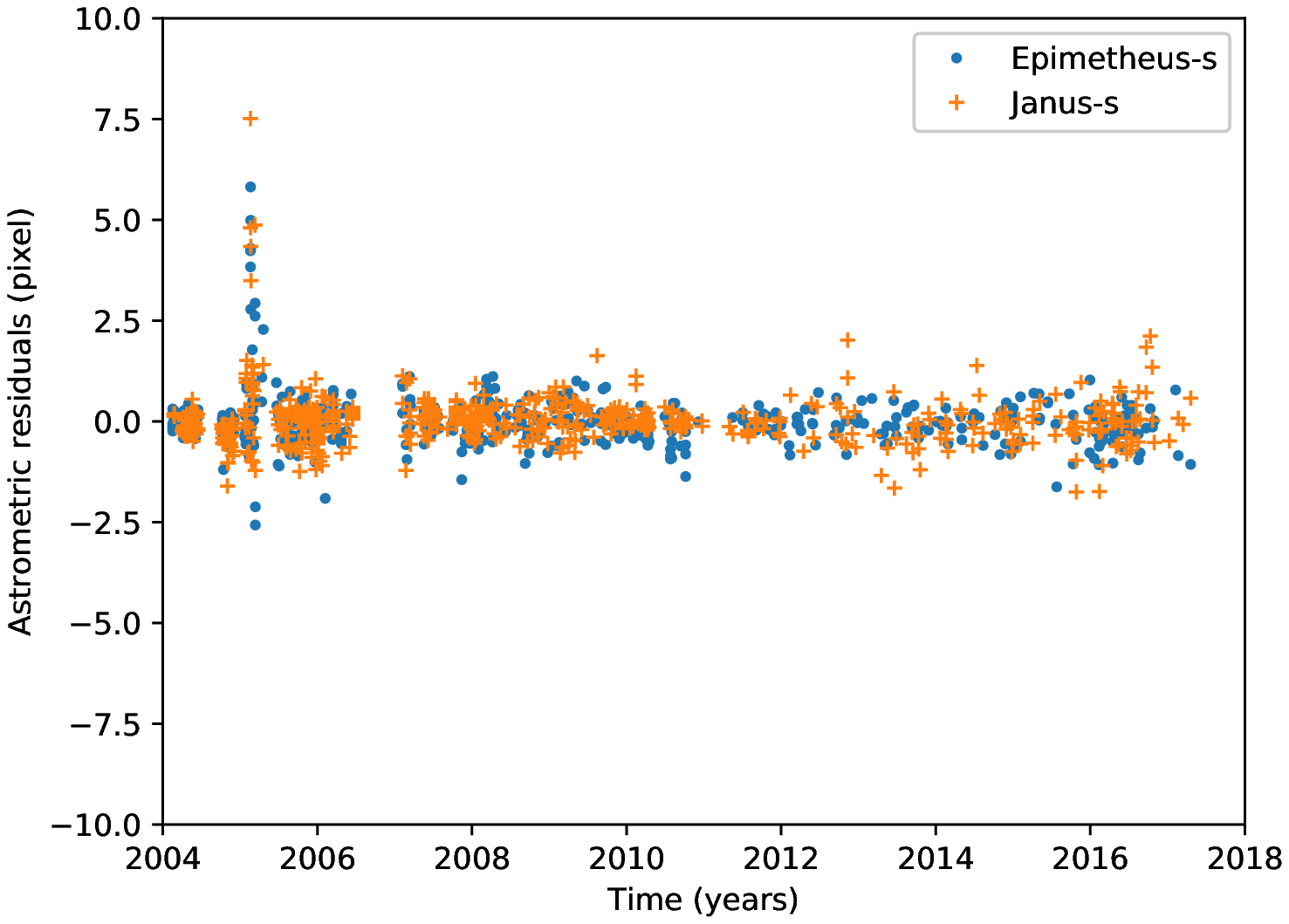} & \includegraphics[width=7.5cm,angle=0]{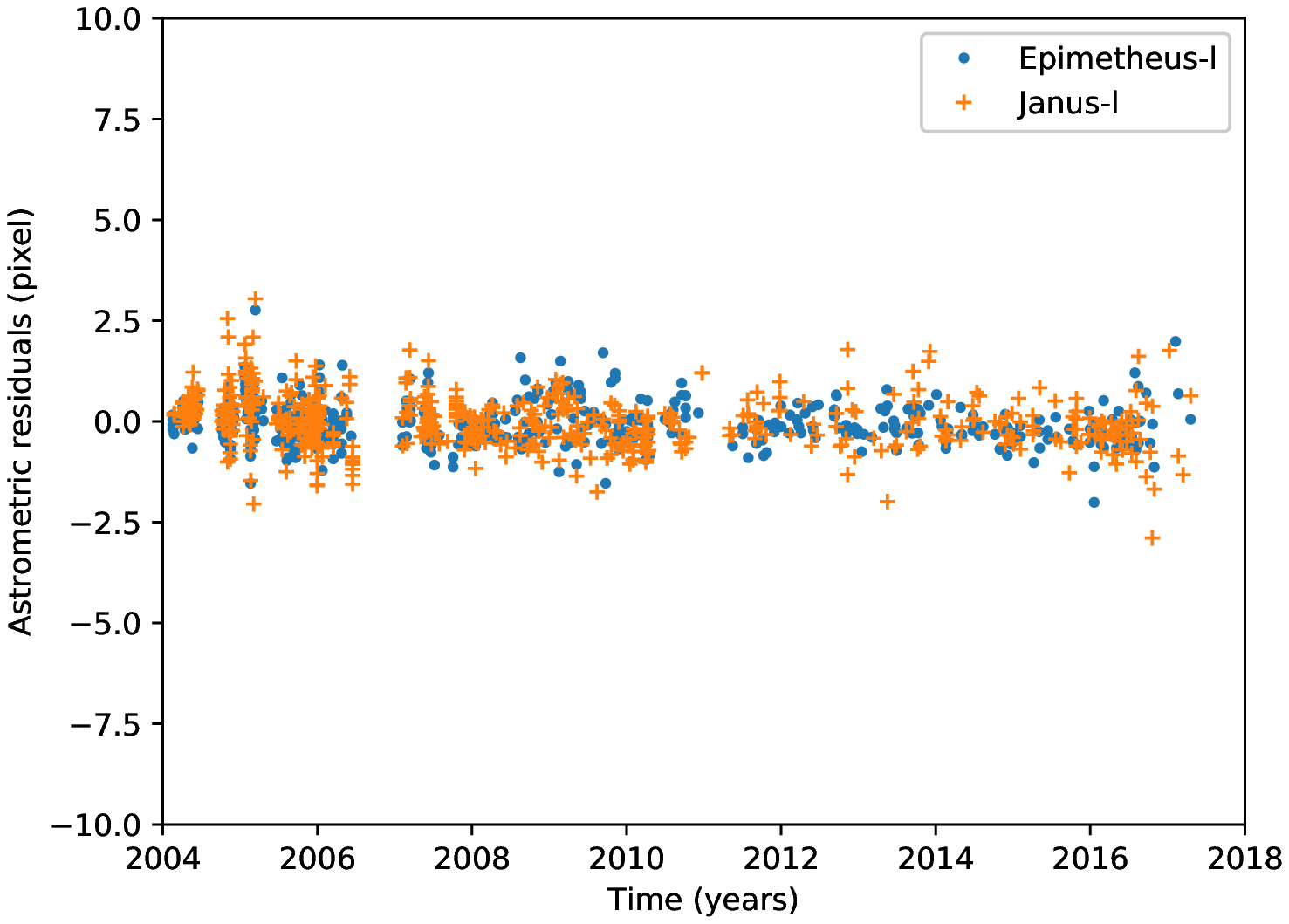}\\
\end{tabular} 
\caption{Astrometric residuals after a fit between the model and the ISS observations for Atlas, Prometheus, Pandora (top), Epimetheus and Janus (bottom) in sample (left) and 
line (right).}\label{fig:resisualsISS}
\end{center} 
\end{figure*}

\begin{figure*} 
\begin{center} 
\begin{tabular}{ll} 
\hspace*{-0.35cm}\includegraphics[width=7.5cm,angle=0]{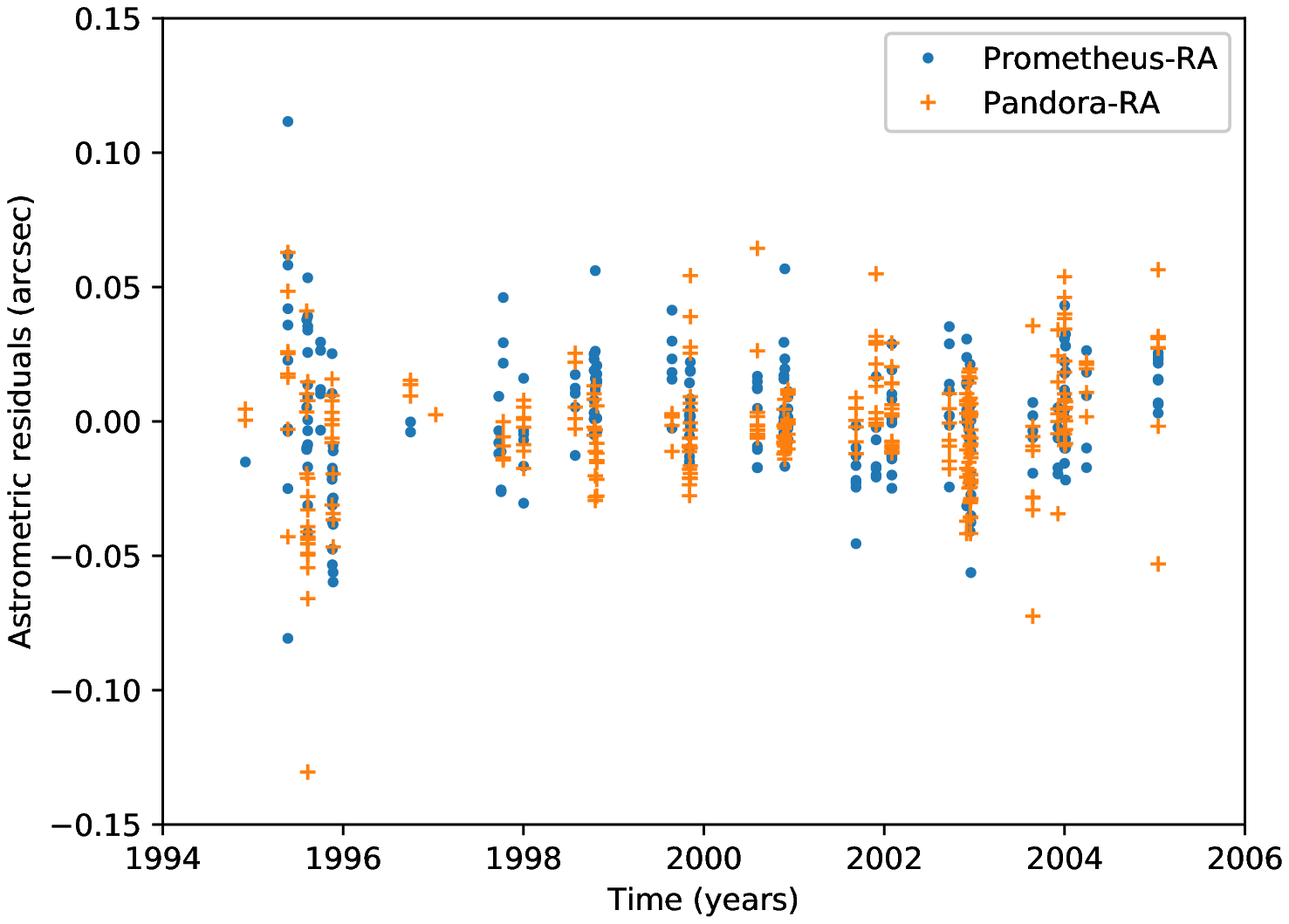} & \includegraphics[width=7.5cm,angle=0]{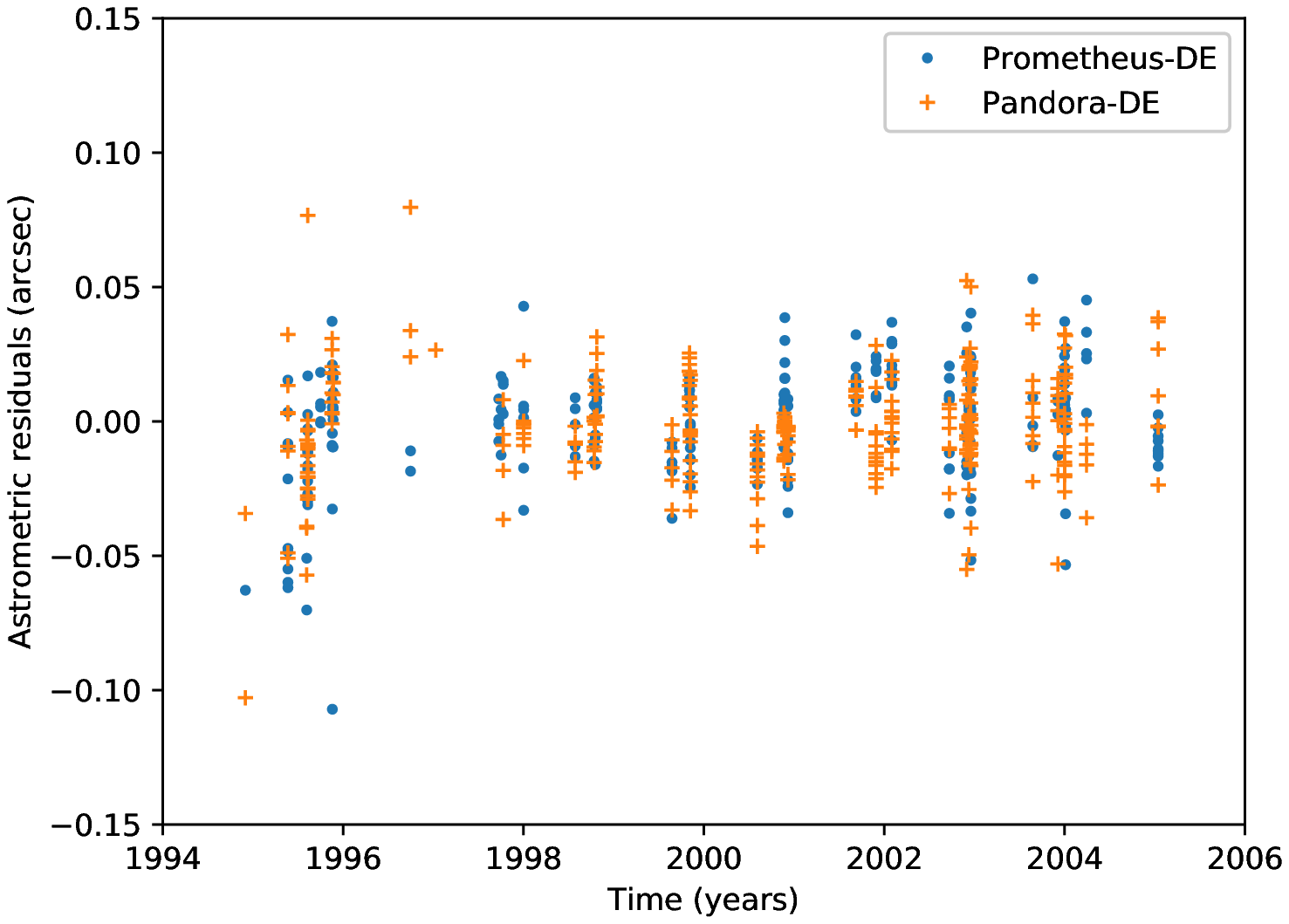}\\
\hspace*{-0.35cm}\includegraphics[width=7.5cm,angle=0]{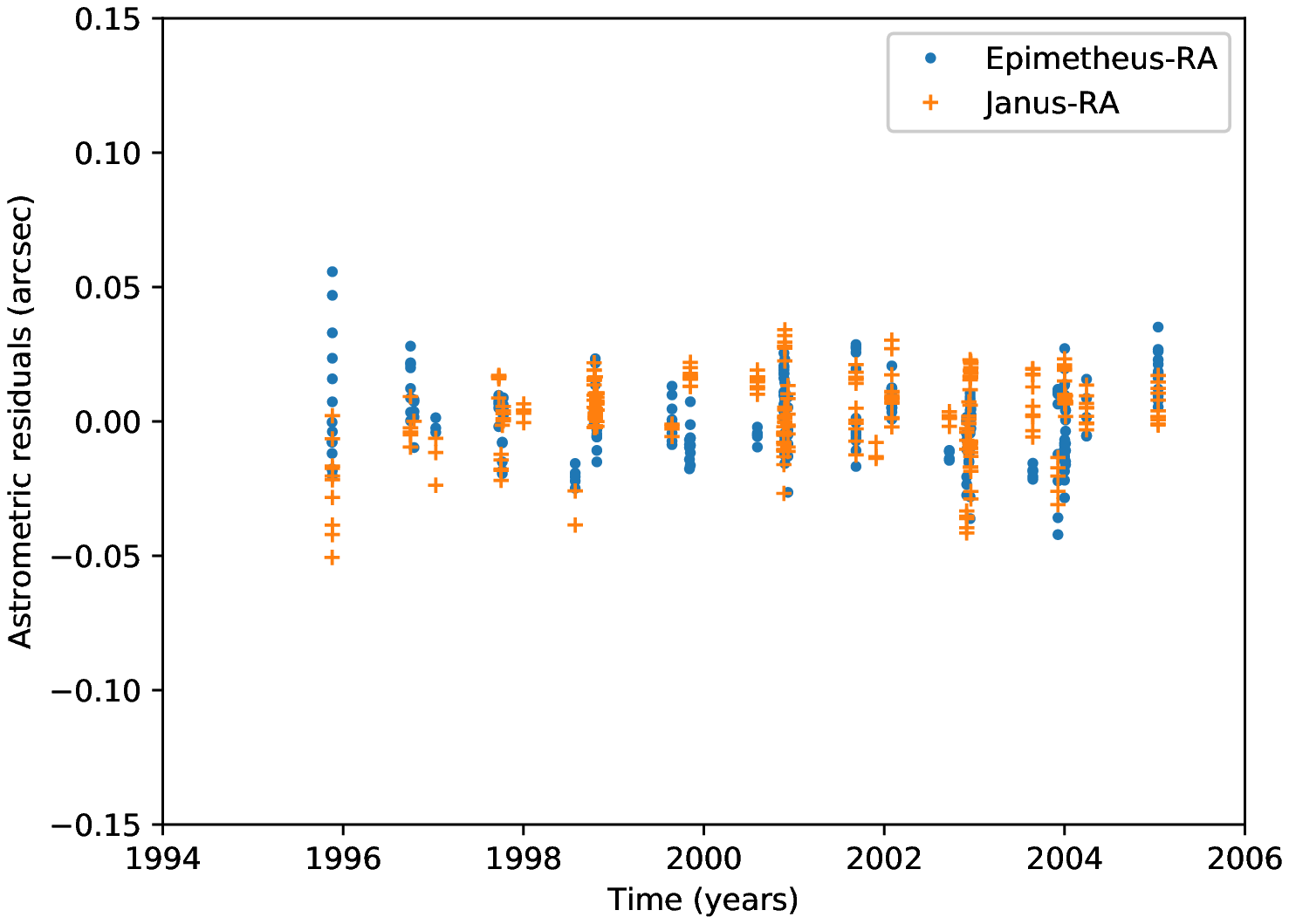} & \includegraphics[width=7.5cm,angle=0]{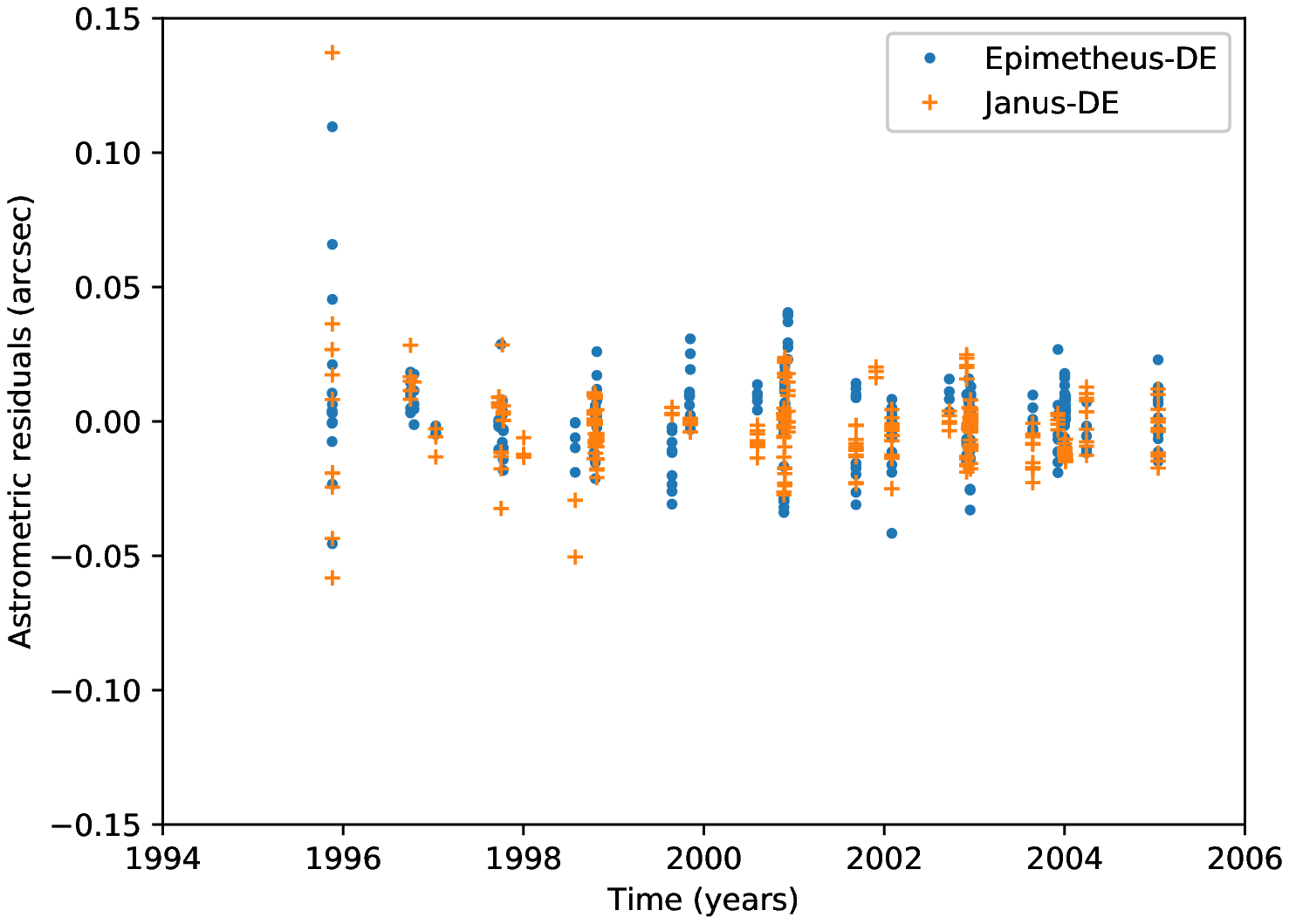}\\
\end{tabular} 
\caption{Astrometric residuals after a fit between the model and the HST observations for Prometheus and Pandora (top), Epimetheus and Janus (bottom) in right ascension (left) and 
declination (right).}\label{fig:resisualsHST}
\end{center} 
\end{figure*}

\begin{figure*} 
\begin{center} 
\begin{tabular}{ll} 
\hspace*{-0.35cm}\includegraphics[width=7.5cm,height=7.5cm,angle=0]{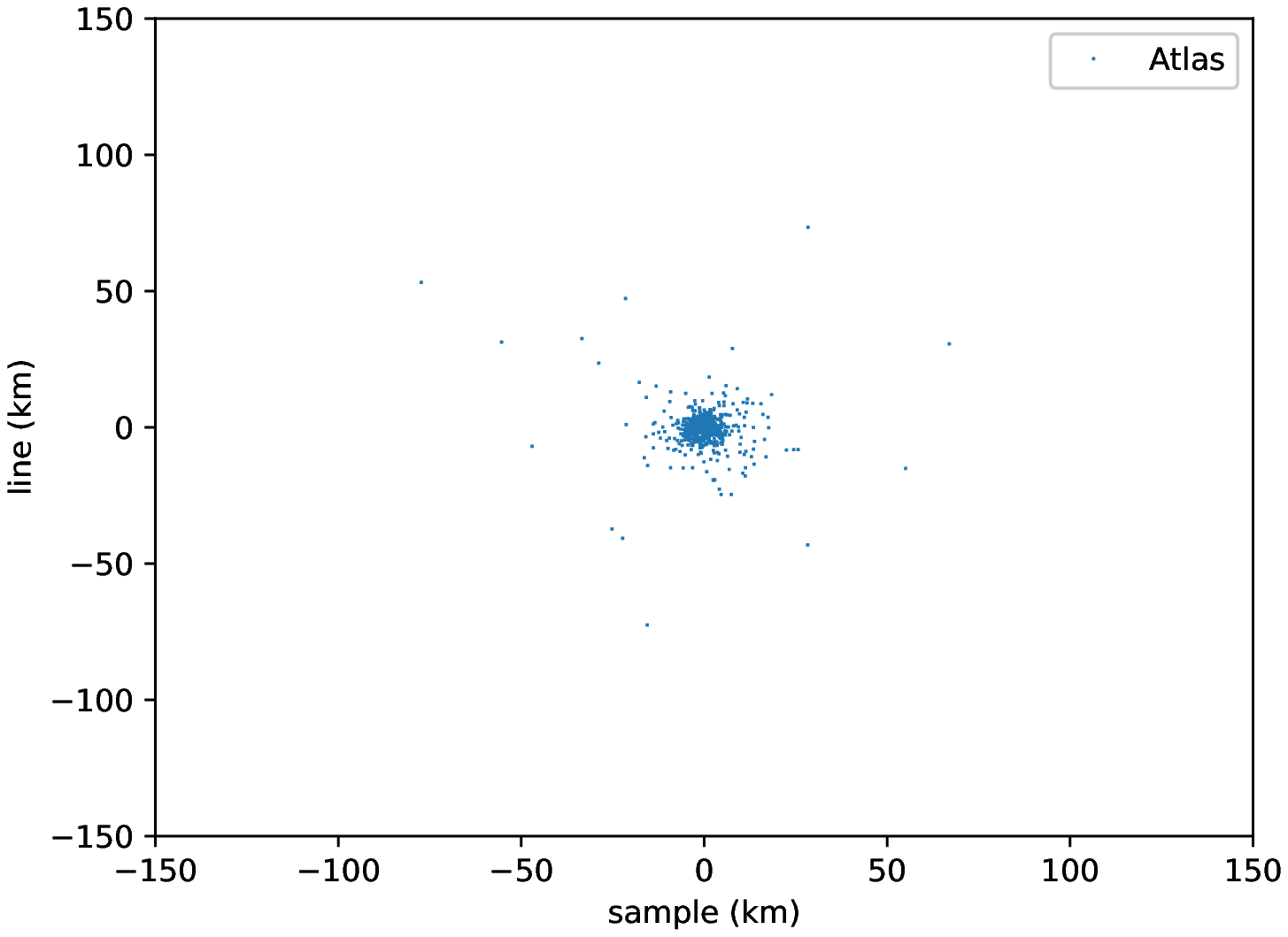} & \includegraphics[width=7.5cm,height=7.5cm,angle=0]{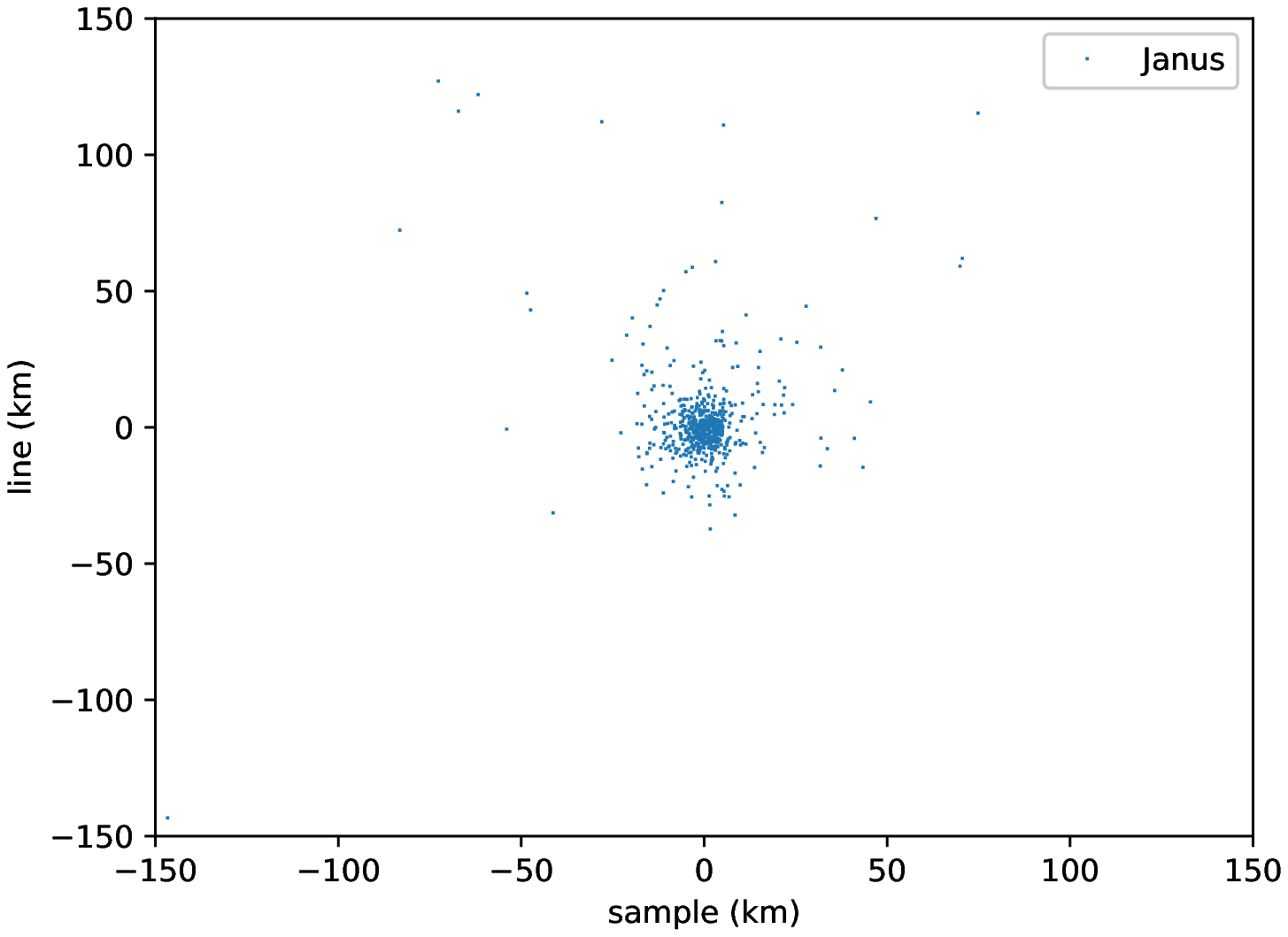}\\
\hspace*{-0.35cm}\includegraphics[width=7.5cm,height=7.5cm,angle=0]{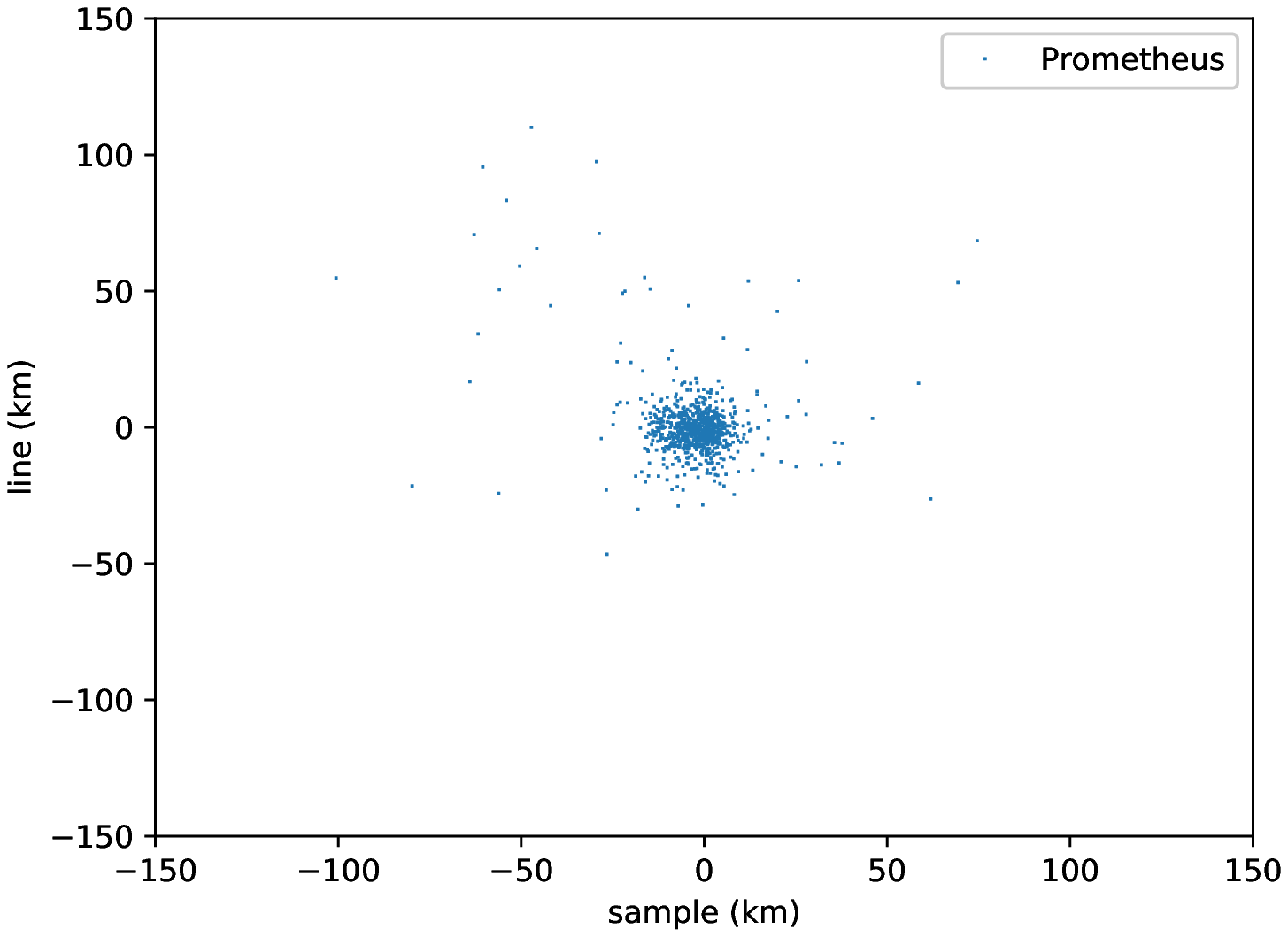} & \includegraphics[width=7.5cm,height=7.5cm,angle=0]{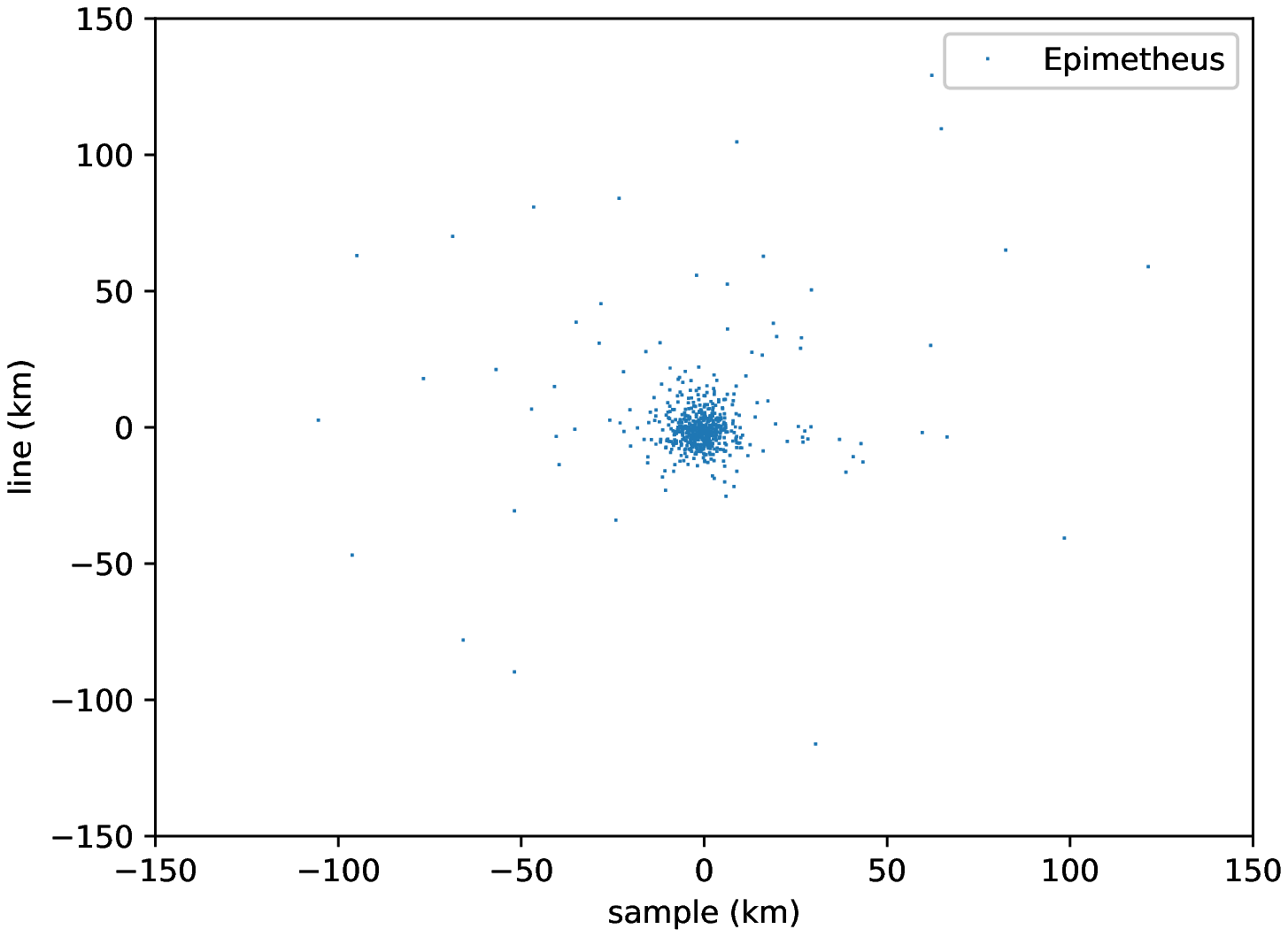}\\
\hspace*{-0.35cm}\includegraphics[width=7.5cm,height=7.5cm,angle=0]{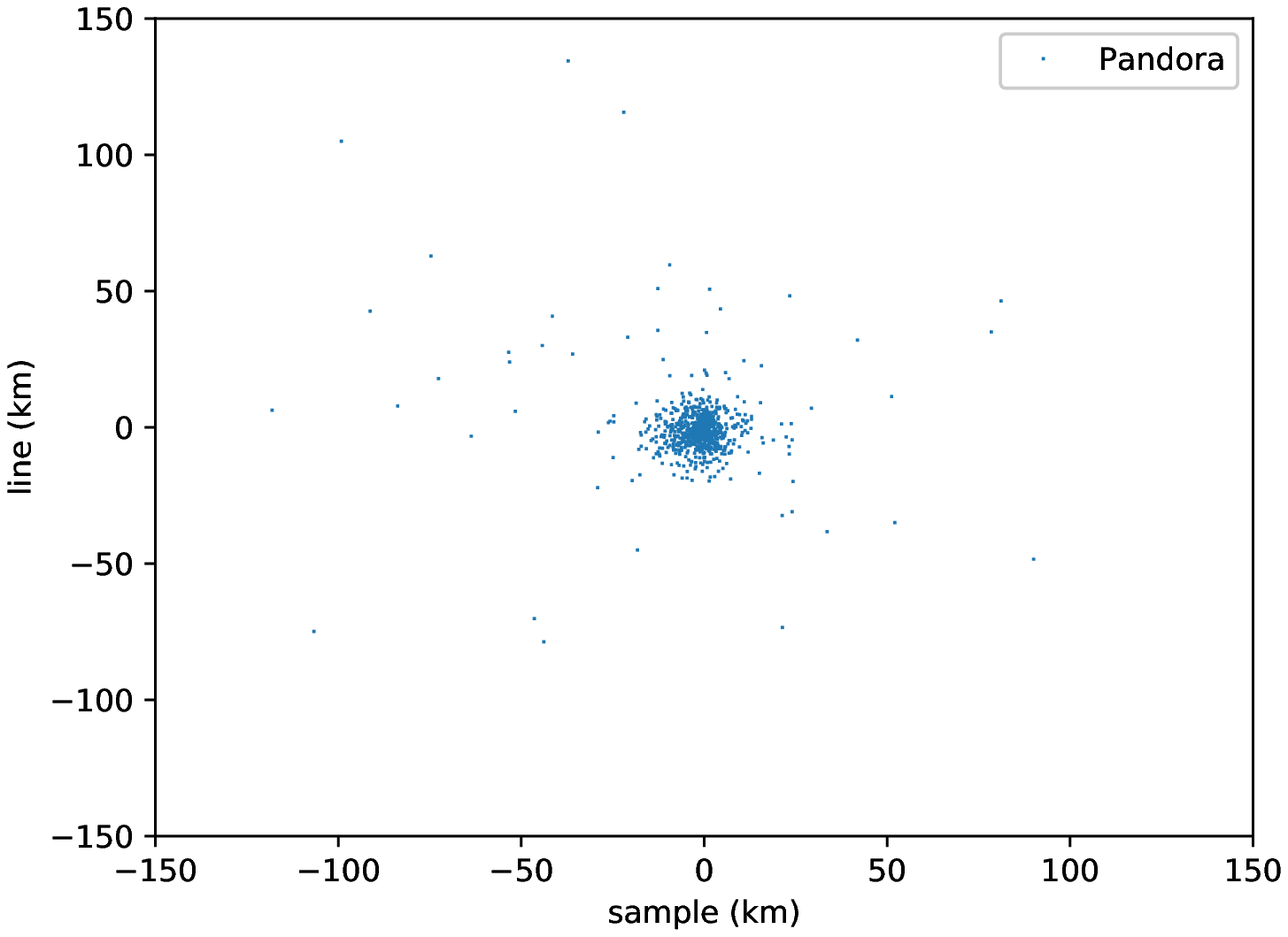} & \\
\end{tabular} 
\caption{Astrometric residuals after a fit between the model and the ISS observations for all moons in sample and line coordinates rescaled in $km$.}\label{fig:resisualsISS2}
\end{center} 
\end{figure*}

As an example, we provide here detailed information for the simulation that introduces both ISS and HST data.
In Figures \ref{fig:resisualsISS}, \ref{fig:resisualsHST}, \ref{fig:resisualsISS2} and Tables \ref{tab:moysig}, \ref{tab:moysig2} we give the astrometric residuals for the five inner moons.
Our residuals are pretty close to the ones obtained from \citet{Coop15}, except for the absence of Solar bias and the use of extra observations.
\newline

\begin{table*} 
\caption{Mean ($\nu$) and standard deviation ($\sigma$) on sample and line in pixel for each satellite (Cassini-ISS data). $N$ is the number of observations by satellite considered and for each coordinate. $\bar{N}$ is the number of observations by satellite when counting successive observation series as one. Observations with residuals higher than 2.5 pixels were discarded. NAC and WAC stand for the Narrow Angle Camera and Wide Angle Camera of the Cassini Imaging Science Subsystem, respectively.}
\begin{center} 
\begin{small}
\begin{tabular}{lrrrrrl} 
\hline
\hline
Observations&$\nu_{sample}\ \ \ $&$\sigma_{sample}\ \ \ $&$\nu_{line}\ \ \ $&$\sigma_{line}\ \ \ $&$N\ \ $&satellite\\
& (pix)$\ \ \  $& (pix)$\ \ \  $& (pix)$\ \ \  $& (pix)$\ \ \  $&  & \\
 \hline
ISS NAC  (centroid fitting) 
prearrival &             0.0139 &             0.2934 &            -0.0802 &             0.3104 &     32,     32 & Atlas\\
&            -0.0714 &             0.2801 &             0.1593 &             0.3129 &     53,     53 & Prometheus\\
&            -0.0458 &             0.2842 &             0.1003 &             0.3034 &     58,     58 &Pandora\\
&            -0.0142 &             0.2205 &             0.1313 &             0.2412 &     54,     54 &Epimetheus\\
&            -0.0222 &             0.2106 &             0.3209 &             0.2754 &     41,     41 &Janus\\
 \hline
ISS NAC  (limb fitting)
 &            -0.1300 &             0.4813 &            -0.0603 &             0.4303 &     192,    194 & Atlas\\
 &            -0.2004 &             0.5879 &            -0.0896 &             0.5991 &     804,    805 & Prometheus\\
 &            -0.1109 &             0.4876 &            -0.1160 &             0.5052 &    718,    725 &Pandora\\
 &            -0.0706 &             0.4553 &            -0.0752 &             0.4946 &     534,    542 &Epimetheus\\
 &            -0.0087 &             0.5106 &            -0.0435 &             0.6382 &    520,    522 &Janus\\
 \hline
ISS WAC      (limb fitting)   
& -1.0290 &             0.0000 &            -1.0054 &             0.0000 &     1,      1 &Janus\\
\hline
ISS NAC  (centroid fitting) 
&             0.0513 &             0.3678 &            -0.0037 &             0.4108 &     425,    425 & Atlas\\
&             0.4541 &             0.7275 &            -0.3418 &             0.0721 &       2,      2 & Prometheus\\
&             0.1602 &             0.4842 &            -0.2783 &             0.1616 &       6,      6 &Pandora\\
&            -0.1417 &             0.7008 &            -0.0524 &             1.0145 &       4,      4 &Epimetheus\\
 \hline
\end{tabular} 
\end{small}
\end{center} 
\label{tab:moysig} 
\end{table*}

\begin{table*} 
\caption{Mean ($\nu$) and standard deviation ($\sigma$) on $\Delta \alpha \cos(\delta)$ and $\Delta \delta$ in arcsec for each satellite (HST data). $N$ is the number of observations by satellite considered and for each coordinate.}
\begin{center} 
\begin{small}
\begin{tabular}{lrrrrrl} 
\hline
\hline
Observations&$\nu_{\alpha\cos{\delta}}\ \ \ $&$\sigma_{\alpha\cos{\delta}}\ \ \ $&$\nu_{\delta}\ \ \ $&$\sigma_{\delta}\ \ \ $&$N\ \ $&satellite\\
& (arcsec)$\ \ \  $& (arcsec)$\ \ \  $& (arcsec)$\ \ \  $& (arcsec)$\ \ \  $&  & \\
\hline
	\citet{Fren03} HST PC                                         
 &            -0.0012 &             0.0355 &            -0.0132 &             0.0277 &     52,     52  & Prometheus\\
 &            -0.0151 &             0.0349 &            -0.0076 &             0.0287 &     44,     44  &Pandora\\
 &             0.0068 &             0.0238 &             0.0137 &             0.0369 &     14,     14  &Epimetheus\\
 &            -0.0240 &             0.0154 &             0.0061 &             0.0524 &     10,     10  &Janus\\
 \hline
	\citet{Fren06} HST PC                       
&             0.0005 &             0.0168 &             0.0015 &             0.0165 &    199,    199  & Prometheus\\
&            -0.0022 &             0.0138 &            -0.0009 &             0.0144 &    182,    182  &Pandora\\
&            -0.0017 &             0.0113 &             0.0004 &             0.0141 &    172,    172  &Epimetheus\\
&             0.0051 &             0.0133 &            -0.0015 &             0.0108 &    180,    180  &Janus\\
 \hline
	\citet{Fren06} HST WF4
 &             0.0230 &             0.0185 &             0.0173 &             0.0160 &      9,      9   & Prometheus\\
 &             0.0190 &             0.0207 &            -0.0168 &             0.0258 &     14,     14  &Pandora\\
 &             0.0144 &             0.0072 &             0.0018 &             0.0161 &     21,     21  &Epimetheus\\
 &             0.0012 &             0.0158 &            -0.0074 &             0.0142 &     31,     31  &Janus\\

 \hline
	\citet{Fren06} HST WF3  
 &             0.0112 &             0.0130 &             0.0226 &             0.0058 &      5,      5   & Prometheus\\
 &            -0.0062 &             0.0207 &             0.0027 &             0.0287 &     15,     15  &Pandora\\
 &             0.0000 &             0.0143 &            -0.0020 &             0.0077 &     19,     19  &Epimetheus\\
 &             0.0088 &             0.0076 &             0.0007 &             0.0067 &     12,     12  &Janus\\
 \hline
	\citet{Fren06} HST WF2               
 &             0.0077 &             0.0278 &             0.0146 &             0.0226 &      3,      3    & Prometheus\\
 &             0.0066 &             0.0345 &             0.0061 &             0.0224 &     26,     26  &Pandora\\
 &            -0.0001 &             0.0190 &             0.0012 &             0.0140 &     33,     33  &Epimetheus\\
 &            -0.0082 &             0.0101 &            -0.0100 &             0.0097 &     28,     28  &Janus\\
\hline
\end{tabular} 
\end{small}
\end{center} 
\label{tab:moysig2} 
\end{table*}

While \citet{Coop15} solved only for the initial state vectors and masses of the inner moons, all our solutions always included the orientation and precession of Saturn's pole plus the $J_2, J_4, J_6$ gravity coefficients. 
Despite the larger number of fitted coefficients, our uncertainties are about the same order of magnitude thanks to the use of better weighting procedure and the removal of the Solar bias. Tables \ref{tab:CI} and \ref{tab:paramphys} provide initial state vectors and physical parameters of interest for the solution that includes HST data. Quoted uncertainties are 1 $\sigma$. It is noteworthy to mention that the orientation and precession of Saturn's pole was recently determined by \citet{Fren17} with much better precision.
\newline

\begin{table*} 
\caption{Solution for the Planetocentric State Vectors in the ICRF at Epoch 2005 January 1 00:00:00.0 TDB (2453371.5 JD). 
Units are $km^3s^{-2}$, $km$ and $km/sec$ for $Gm$s, positions and velocities, respectively.}
\begin{center} 
\begin{small}
\begin{tabular}{lrrrrrr} 
\hline
\hline
Atlas  &&&\\
$Gm$&   3.729516555394247E-4 $\ \pm\ $ 8.3E-7 & & \\
          $x,y,z$                  &   137001.867291721      $\ \pm\ $ 0.14 &    4781.60971003271     $\ \pm\ $ 1.2 &    -12140.3481577703        $\ \pm\ $  0.31\\
 $\dot{x},\dot{y},\dot{z}$  & -0.700351459623150     $\ \pm\ $ 1.5E-4&     16.5882698930507    $\ \pm\ $ 1.7E-5&     -1.16305098637571        $\ \pm\ $ 4.2E-5 \\

Prometheus  &&&\\
$Gm$&   1.066670940945999E-2 $\ \pm\ $ 3.0E-6 & & \\
          $x,y,z$                  &  -31658.8799258814     $\ \pm\ $ 0.86 &     135240.714417874    $\ \pm\ $ 0.24 &     -7245.73352485315        $\ \pm\ $ 0.38  \\
 $\dot{x},\dot{y},\dot{z}$  &  -16.0692801273308     $\ \pm\ $ 2.9E-5&    -3.68640166175008    $\ \pm\ $ 1.0E-4&      1.65619629017423        $\ \pm\ $ 3.9E-5 \\

Pandora &&&\\
$Gm$&   9.097099552691643E-3 $\ \pm\ $ 4.7E-6 & & \\
          $x,y,z$                  &  -116624.718116507     $\ \pm\ $ 0.58 &     80455.0482120157     $\ \pm\ $ 0.81 &     3980.43591260067        $\ \pm\ $ 0.38 \\
 $\dot{x},\dot{y},\dot{z}$  &  -9.26910715040568    $\ \pm\ $ 9.0E-5&     -13.4034429873202    $\ \pm\ $ 6.4E-5&      1.78602159392975        $\ \pm\ $ 3.4E-5 \\

Epimetheus  &&&\\
$Gm$&   3.507640402593114E-2 $\ \pm\ $ 4.2E-6 & & \\
          $x,y,z$                  &  -141847.308320134     $\ \pm\ $ 0.32 &     48917.8663153782     $\ \pm\ $ 0.98 &     8200.72245869048        $\ \pm\ $  0.43 \\
 $\dot{x},\dot{y},\dot{z}$  &  -5.00404958200978    $\ \pm\ $ 8.4E-5 &     -15.0870443324898     $\ \pm\ $ 4.4E-5 &     1.63209435848272        $\ \pm\ $ 3.9E-5 \\

Janus  &&&\\
$Gm$&    0.126390571242701 $\ \pm\ $ 1.6E-5 & & \\
          $x,y,z$                  &  -32298.9036418660    $\ \pm\ $ 0.82 &     -148325.820102700    $\ \pm\ $ 0.30 &      14126.3843856947        $\ \pm\ $ 0.46 \\
 $\dot{x},\dot{y},\dot{z}$  &   15.3284304419310    $\ \pm\ $ 3.0E-5 &     -3.46497045237302    $\ \pm\ $ 6.1E-5 &     -1.07788018919314        $\ \pm\ $ 4.5E-5 \\
\hline
\end{tabular} 
\end{small}
\end{center} 
\label{tab:CI} 
\end{table*}

\begin{table*} 
\caption{Solution for the primary's gravity field and polar orientation. T denotes Julian century. The epoch is J2000.}
\begin{center} 
\begin{small}
\begin{tabular}{lrr} 
\hline
\hline
 $R$   (km)   &  60330.0  &\\
 $J_2$    &   1.627545066665849E-002  $\ \pm\ $& 1.30E-005\\
 $J_4$    &  -9.630492172453784E-004   $\ \pm\ $& 5.99E-005\\
 $J_6$     &  1.250890032746516E-004   $\ \pm\ $& 9.75E-005\\
 $\alpha$  (deg)& 40.583475082321  $\ \pm\ $& 1.28E-3\\
 $\dot{\alpha}$  (deg/T) & -5.082364097807413E-002  $\ \pm\ $& 0.012E0\\
 $\delta$  (deg)& 83.53783607375815   $\ \pm\ $& 1.34E-4\\
 $\dot{\delta}$  (deg/T)& -5.709661021904670E-003  $\ \pm\ $& 0.0012E0\\
\hline
\end{tabular} 
\end{small}
\end{center} 
\label{tab:paramphys} 
\end{table*}

In Table \ref{tab:sigmaCompMass} we provide our estimations of the moon's densities. These are pretty close to the ones of \citet{Thom13} and the ones that can be derived from the masses estimated in \citet{Coop15}.
For high reliability, we provide error bars obtained from the envelope of $3 \sigma$ uncertainty associated with our seven main solutions (recall that the three others were performed for test purposes only).
We do confirm the significant leap in density between Atlas (0.39 $g/cm^3$), the pair Prometheus and Pandora (0.48 $g/cm^3$), and the pair Janus and Epimetheus (0.63 $g/cm^3$). Interestingly the two pair satellites have in both cases indistinguishable densities, despite the small error bars. Such features must be a consequence of their formation process.
\newline

\begin{table*} 
	\caption{Estimation of the moons' Gm ($km^3/s^2$). Volume is given in $km^3$ from \citet{Thom13}. Density is given in $g/cm^3$.}
\begin{center} 
\begin{small}
\begin{tabular}{lrrrrr} 
\hline
\hline
Simulation&   Atlas & Prometheus &  Pandora & Epimetheus & Janus  \\ 
Cassini ($J_6$ not& 3.7231E-4$\pm$7.6E-7& 1.06615E-2$\pm$3.0E-6 & 9.0855E-3$\pm$4.5E-6  & 3.50793E-2$\pm$4.2E-6 & 0.126401$\pm$1.6E-5 \\
constrained)& &&&&\\
Cassini [2004-2010]&    3.694E-4$\pm$1.4E-6 &  1.06625E-2$\pm$7.1E-6  &  9.134E-3$\pm$1.4E-5  & 3.50854E-2$\pm$5.2E-6  & 0.126425$\pm$1.9E-5  \\
Cassini [2011-2017]& 3.769E-4$\pm$1.7E-6 &  1.06454E-2$\pm$9.3E-6 &  9.162E-3$\pm$2.0E-5  & 3.50826E-2$\pm$8.0E-6  & 0.12641$\pm$3.0E-5 \\
\hline
Cassini& 3.721E-4$\pm$7.6E-7 & 1.06611E-2$\pm$3.0E-6  & 9.0847E-3$\pm$4.5E-6  &  3.50788E-2$\pm$4.2E-6  & 0.126398$\pm$1.6E-5  \\
Cassini (sat359l eph.)&  3.7171E-4$\pm$7.5E-7 & 1.06576E-2$\pm$3.0E-6  & 9.0788E-3$\pm$4.5E-6   & 3.50799E-2$\pm$4.2E-6 & 0.126403$\pm$1.6E-5  \\
Cassini (sat375 eph.)& 3.7178E-4$\pm$7.6E-7 & 1.06587E-2$\pm$3.0E-6 & 9.0793E-3$\pm$4.5E-6  & 3.50795E-2$\pm$4.2E-6 & 0.126402$\pm$1.6E-5  \\  
Cassini + HST& 3.7295E-4$\pm$8.3E-7  &   1.06667E-2$\pm$3.0E-6  &   9.0971E-3$\pm$4.7E-6     & 3.50764E-2$\pm$4.2E-6   & 0.126391$\pm$1.6E-5     \\  
Cassini (incl. $J_3$, $J_5$)& 3.7220E-4$\pm$7.6E-7 & 1.06606E-2$\pm$3.0E-6 & 9.0848E-3$\pm$4.5E-6 & 3.50796E-2$\pm$4.2E-6  & 0.126401$\pm$1.6E-5 \\  
Cassini (no nutations)& 3.7196E-4$\pm$7.6E-7 & 1.06609E-2$\pm$3.0E-6 & 9.0833E-3$\pm$4.5E-6  & 3.50779E-2$\pm$4.2E-6 & 0.126396$\pm$1.6E-5 \\  
Cassini (Pallene)&  3.7199E-4$\pm$7.5E-7 & 1.06609E-2$\pm$3.0E-6 & 9.0849E-3$\pm$4.5E-6 &  3.50777E-2$\pm$4.2E-6 & 0.12639$\pm$1.6E-5 \\  
\hline
Global $Gm$ solutions&  (3.73$\pm$0.06)$\times10^{-4}$ & (1.066$\pm$0.001)$\times 10^{-2}$ & (9.09$\pm$0.02)$\times 10^{-3}$ & (3.508$\pm$0.001)$\times 10^{-2}$ &  0.12640$\pm$0.00005 \\
Volume&  14368$\pm$1487 & 335131$\pm$17780 & 281015$\pm$18658 &  825589$\pm$40393 &  3022853$\pm$55050  \\
Density&  0.389$\pm$0.041 & 0.477$\pm$0.025 & 0.485$\pm$0.032 & 0.637$\pm$0.031 &  0.627$\pm$0.011 \\
\hline
$Gm$ from& (3.84$\pm$0.01)$\times10^{-4}$&(1.0677$\pm$0.0006)$\times10^{-2}$&(9.133$\pm$0.009)$\times10^{-3}$& (3.5110$\pm$0.0009)$\times10^{-2}$&0.12651$\pm$0.00003\\
Cooper et al. (2015)&&&&&\\
\hline
\end{tabular} 
\end{small}
\end{center} 
\label{tab:sigmaCompMass} 
\end{table*}

In Table \ref{tab:sigmaComp} we provide our estimations of the moon's physical librations. We also attempted a fit with HST data only, to emphasize the precision of ISS data.
For high reliability, we provide error bars obtained from the envelope of $5 \sigma$ uncertainty associated with our seven main solutions (recall again that the three others were performed for test purposes only).
In particular, we did not choose $3 \sigma$ uncertainties since we had neglected higher order in the satellite gravity fields (representing a departure from a purely ellipsoidal shape for an homogeneous body).
In addition, since we assumed specific $J_2$ and $C_{22}$ values for the moons derived from homogeneity assumption, we added the volume uncertainty of each moon in our error bars of the physical librations $\gamma$ (last line of Table \ref{tab:sigmaComp}). This was done\footnote{Volume uncertainty was approached by another way, also. The second method consisted in adding as free parameters in the fit the satellites' $J_2$ and $C_{22}$, but constrained to their uncertainty deduced from volume uncertainty and homogeneous assumption. Both methods provided the same results within less than 8\% (typically 5\% most of the time), showing that the periapsis drift is, indeed, the essential part of the orbital signal coming from physical libration.} from differentiation of equation \ref{eq:libr}.
\newline

\begin{table*} 
\caption{Estimation of tidal and physical libration amplitudes. For practical reasons, in our code we actually fit ${\cal B}$ (see equation \ref{eq:librationNOE}). One can get back to the amplitude of the tidal libration using Eq. \ref{eq:librationNOE}. Physical libration amplitudes (in degrees) are given in the last row of the Table.}
\begin{center} 
\begin{small}
\begin{tabular}{lrrrr} 
\hline
\hline
 Simulation   &   ${\cal B}$ Prometheus &  ${\cal B}$ Pandora & ${\cal B}$ Epimetheus &  ${\cal B}$ Janus  \\ 
  HST    &  - $\pm$ 142. &  - $\pm$ 326. &  - $\pm$ 53. &  - $\pm$ 27. \\
  HST  (masses+libr. only) &  - $\pm$ 25. &  - $\pm$ 38. &  - $\pm$ 22. &  - $\pm$ 15. \\
  Cassini ($J_6$ not constrained) & 0.0 $\pm$ 0.8 & 36.2 $\pm$ 4.2 & 26.6 $\pm$ 1.3 & 2.5 $\pm$ 0.7 \\
  Cassini [2004-2010] &   1.0 $\pm$ 1.8 & 35.8  $\pm$  4.6 & 23.5 $\pm$   2.8 & 6.4$\pm$    1.6 \\
  Cassini [2011-2017] &  0.6 $\pm$ 1.4 &  46.1 $\pm$ 3.4 &  22.1 $\pm$ 5.3 &  1.8 $\pm$ 3.2 \\
  \hline
  Cassini & 1.3 $\pm$ 0.3 & 42.9 $\pm$ 1.3 & 26.6 $\pm$ 1.3 & 2.5 $\pm$ 0.7 \\
  Cassini (sat359l eph.) &   1.3 $\pm$ 0.3 &  42.5 $\pm$ 1.3 &   26.8 $\pm$ 1.3 &  2.4 $\pm$ 0.7\\
  Cassini (sat375 eph.) &  1.3 $\pm$ 0.3 &  42.7 $\pm$ 1.3 &  26.7 $\pm$ 1.3 &  2.4 $\pm$ 0.7 \\  
  Cassini + HST &  1.5 $\pm$ 0.3 & 43.5 $\pm$ 1.3 &  26.0 $\pm$ 1.3 &  3.2 $\pm$ 0.7 \\
  Cassini (incl. $J_3$, $J_5$) &  1.3  $\pm$ 0.3 &  42.7 $\pm$ 1.3 & 25.8 $\pm$ 1.3 &  2.4 $\pm$ 0.7 \\  
  Cassini (no nutations) & 1.3 $\pm$ 0.3 & 43.2 $\pm$ 1.3 & 26.5 $\pm$ 1.3 & 2.4 $\pm$ 0.7 \\
  Cassini (Pallene) & 0.9 $\pm$ 0.2 & 40.9 $\pm$ 0.9 & 27.2 $\pm$ 1.3 & 2.7 $\pm$ 0.7 \\  
\hline
  Global solution & 1.55 $\pm$ 1.45 & 43.2 $\pm$ 6.8 &  26.3 $\pm$ 7.0 & 2.8 $\pm$ 3.9 \\
\hline
  Physical libration $\cal A$ (deg)& +0.06 $\pm$ 0.37 &  -9.9 $\pm$ 7.4 & -13.5  $\pm$ 15.8 &  -0.31 $\pm$ 1.62 \\
  including vol. uncertainty &&&&\\
\hline
\end{tabular} 
\end{small}
\end{center} 
\label{tab:sigmaComp} 
\end{table*}

We can see from Table \ref{tab:sigmaComp} that while the physical libration of most moons can be constrained with astrometry (in the case of Atlas, the signature is too small), Pandora's physical libration is the only one that is distinguished from zero. 
Both the Epimetheus and Janus values here are in agreement with the homogeneous assumption, within the uncertainty of the measurements. On the other hand, a significant departure is found for Prometheus and to a less extent Pandora. For this particular case, such a disagreement might arise from the departure from triaxiality of Pandora's shape. In particular, Pandora's proper frequency is not too far from the forced one. Hence, a small error in the computation of the Pandora's shape may imply large variations in the expected forced libration amplitude.
This issue was actually solved by \citet{Tisc09} in the case of Epimetheus, by computing the moments of inertia directly from 3-D shape modeling. A similar approach could elucidate the origin of the libration disagreement for Pandora. The case of Prometheus is more problematic since its free libration is expected to be rather different from the forced one. Possible explanations for this are discussed in section \ref{sec:Disc}.

\subsection{Introducing Epimetheus' libration in the fit, obtained directly from image measurements}\label{sec:Tisc}

Since our measurements use the orbital motion of the moons, instead of direct measurement of their rotation, we tried in a last attempt to introduce the results of \citet{Tisc09} as an extra constraint in one of our solutions. In practice, we constrained Epimetheus' physical libration to remain within 1.2 $\deg$ ($1 \sigma$) around -5.9  $\deg$ and, as a consequence, added as an extra free parameter the moon's gravity coefficient $C_{22}$. We obtained after a few iterations the value ($3 \sigma$ uncertainty): $C_{22}=(3.0 \pm 1.7)\times 10^{-2}$. Now adding the volume $3 \sigma$ uncertainty in the use of $J_2$, we finally get: $C_{22}=(3.0 \pm 1.8)\times 10^{-2}$. 
Hence, we see that one may benefit from both approaches (libration direct measurement and astrometry) to constrain better the interior of moons. 
We discuss this in more detail in the next section.

\section{Discussion}\label{sec:Disc}

\begin{figure*} 
\begin{center} 
\hspace*{-0.35cm}\includegraphics[width=15cm,angle=0]{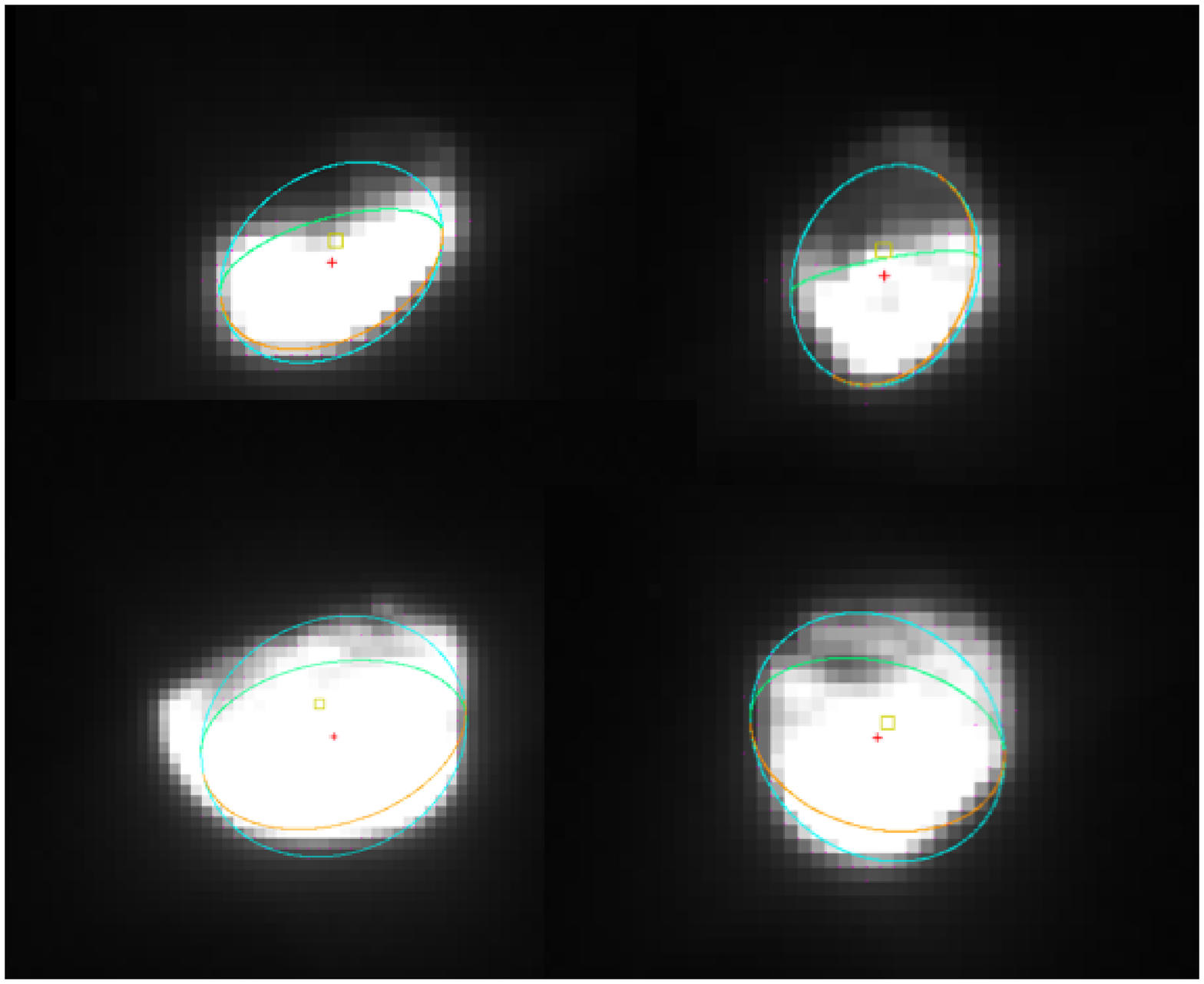} \\
	\caption{A zoom on Prometheus (top left), Pandora (top right), Janus (bottom left) and Epimetheus (bottom right) moon. The orange square shows each predicted positions based on the ephemeris. The red cross shows the measured position based on limb-fitting using the Caviar software \citep{Coop18}. Clearly a more complex shape model is required to benefit fully from the high accuracy of ISS data.}\label{fig:CaviarImages}
\end{center} 
\end{figure*}

Our results demonstrate clearly that the physical librations of at least some of Saturn's small moons could be constrained by monitoring their orbital motion. Nevertheless, we still face two significant issues for a proper conclusion on the interior of these moons. The first problem arises from the significant error bars. The source of this is three fold: i) volume uncertainty; ii) astrometric error measurement; iii) modeling error. While the volume uncertainty is a significant issue for Epimetheus and Pandora, because of the proximity of a resonance between the proper and forced frequencies of these two moons, the astrometric error concerns all five moons. A comparison with astrometric residuals of the main moons \citep{Lain17} shows typically that the inner moons have larger residuals by a factor 2 to 3. Hence, it makes sense to look at the measurement procedure in more detail. As was said in Section \ref{sec:obs}, the Caviar software uses an ellipsoidal shape for fitting the center of figure of the moons. Since the inner moons have strongly distorted shapes, it is expected that a significant part of the astrometric error may come from here. In Figure \ref{fig:CaviarImages} we show a few examples of center of figure fitting with Caviar for the small moons. The use of an ellipsoid as an approximation of the moon's shape clearly shows its limits. Hence, it is likely that the introduction of Digital Terrain Models (DTM) for fitting the shape of such distorted moons will allow residuals to be obtained that are at least two times smaller. Last, a modeling error is still possible. In particular, we obtained an unrealistic value for Saturn's $J_6$ when it was not constrained in the fit. While the exact source of such a bias is uncertain at the moment, the possibility remains that non-zonal harmonics or a temporal variation in the gravity field is affecting the dynamics of the inner moons. Under such circumstances, Prometheus and Pandora should be the most affected by this mismodeling, since they evolve closer to Saturn. This could be the best current explanation for the disagreement between the expected and fitted amplitudes of their physical libration.
\newline

Another issue associated with our method stands on our assumption that the moons are homogeneous. While any significant departure of the fitted physical libration values from an assumption of homogeneity would definitely indicate an inhomogeneous interior, the reverse is not necessarily true. In particular, there can always exist an inhomogenous interior, implying inhomogeneous values for $J_2, C_{22}$ and $\cal{A}$ that will eventually compensate to give a drift of the periapsis (see eq. \ref{eq:libr}) in agreement with a homogeneous model. Such a problem could potentially be mitigated if the gravity coefficient of the moons could be constrained directly from their secular effect on the orbits. In practice, such drifts are strongly correlated to the gravity harmonics of the primary itself. Fortunately, the Saturn system has a large number of moons. In particular, it is not impossible that adding the main moons into a global fit may help in solving for some of the gravity coefficients of the moons. Such work is nevertheless outside the scope of this paper and may be investigated in a further publication.
\newline

It is noteworthy that direct measurement of the rotation has a similar caveat. In principle, observing an amplitude of physical libration in agreement with a homogeneous expectation does not prove the body to be homogeneous. In common with astrometry, one needs to quantify the three moments of inertia $A, B$ and $C$ from observations. This means having three observed quantities, each one giving extra information on the moment of inertia. This can be achieved by quantifying $J_2, C_{22}$ and $\cal A$ from astrometry, or $\cal A$, obliquity and precession rate from direct observation of the rotation. In practice this sounds very challenging. Fortunately, a coupled approach can be used by considering simultaneously direct measurement and astrometry all together. Hence, one might get directly $\cal A$ from observations leaving $J_2$ and $C_{22}$ to astrometry (as done by \citet{Jaco14}. Any other combination will work too. Due to limitations in space mission design, such a coupled approach may be the most rewarding.

\section{Conclusion}

In this work we studied in detail $Cassini$ ISS astrometric data with emphasis on Gaussian behavior.  We showed that the well known solar bias issue can be treated essentially by a simple procedure, with not much effort.
Then, we gave an update on the mass and densities of the inner moons, that fits perfectly former estimations. Last, we tried quantifying the physical libration of these small Saturnian moons. Our results are in agreement with a fully homogeneous interior for Janus and Epimetheus. On the other hand, we found a disagreement for Prometheus and to a less extent Pandora, suggesting a possible dynamical mismodeling. 
We showed that a significant improvement may be expected by reducing the data again using DTM's, as well as by adding the main moons in a global inversion.
In a last attempt, we showed how rotation measurements can be introduced in a synergistic way with astrometric measurements of centers of mass to increase the feedback on interior mass distributions. Such synergy may be useful for further space missions.

\appendix

\section{Solar direction bias}\label{app:1}

\citet{Coop14} demonstrated how factors related to the viewing geometry cause a bias in the individual positions of the order of 0.28 pixel, that becomes systematic across the dataset as a whole.
Such work showed that the solar phase was an issue. Unfortunately, the authors did not provide a clear method for removing properly such a bias. Moreover, such work was done for the main moons, whose shapes are close to spheres. Hence, it is likely that a larger bias may appear on the astrometric data of the small moons.
Here we quantified the Solar direction bias for the small moons, and tried to check its dependency as a function of different geometric factors, like distance and phase angle. 
Then we provided an easy and reasonable way to get rid of the Solar bias issue. The Gaussian behavior of the astrometric data after our treatment is assessed in the next appendix.
\newline

\begin{figure*} 
\begin{center} 
\begin{tabular}{ll} 
\hspace*{-0.35cm}\includegraphics[width=7.5cm,height=7.5cm,angle=0]{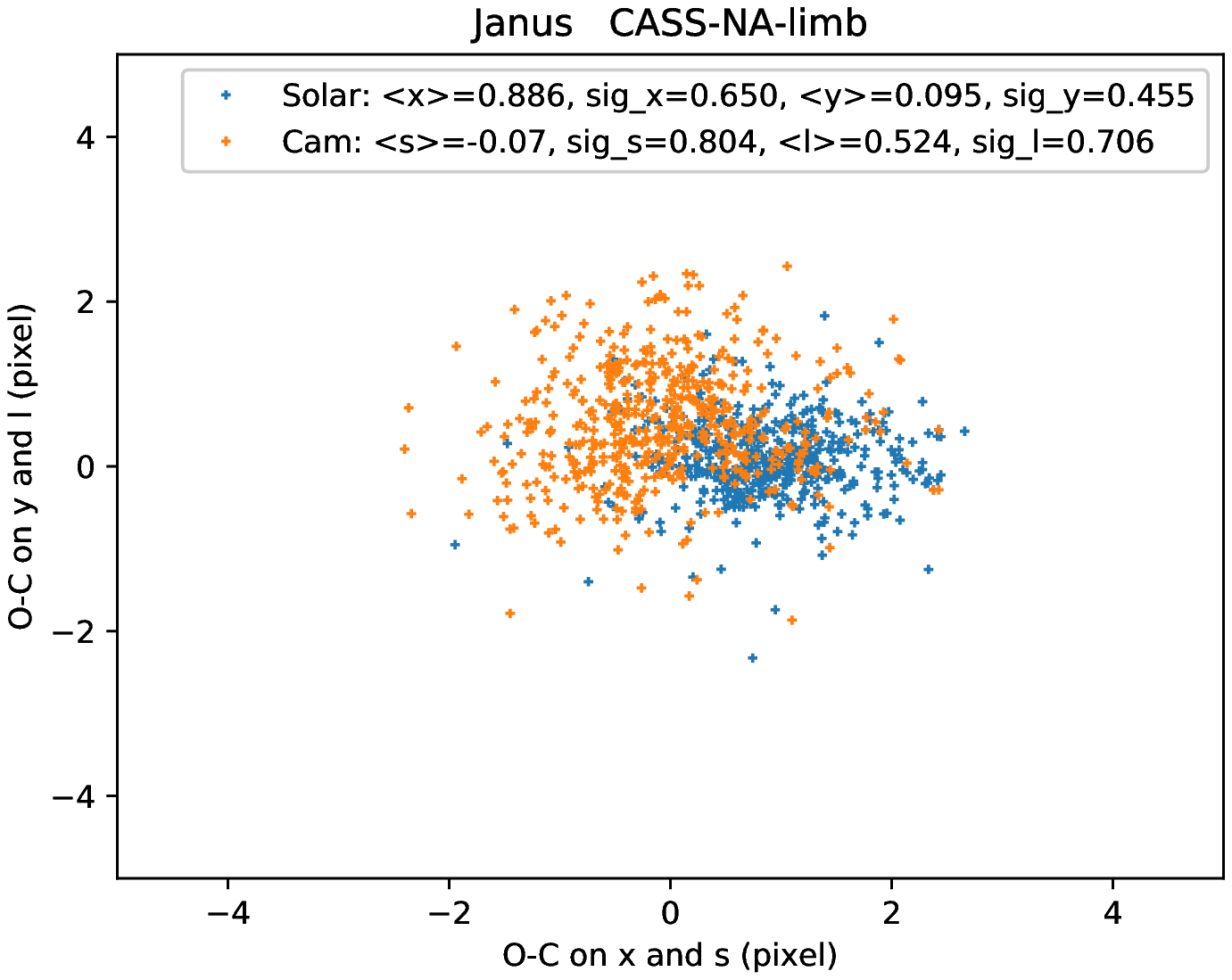} & \includegraphics[width=7.5cm,height=7.5cm,angle=0]{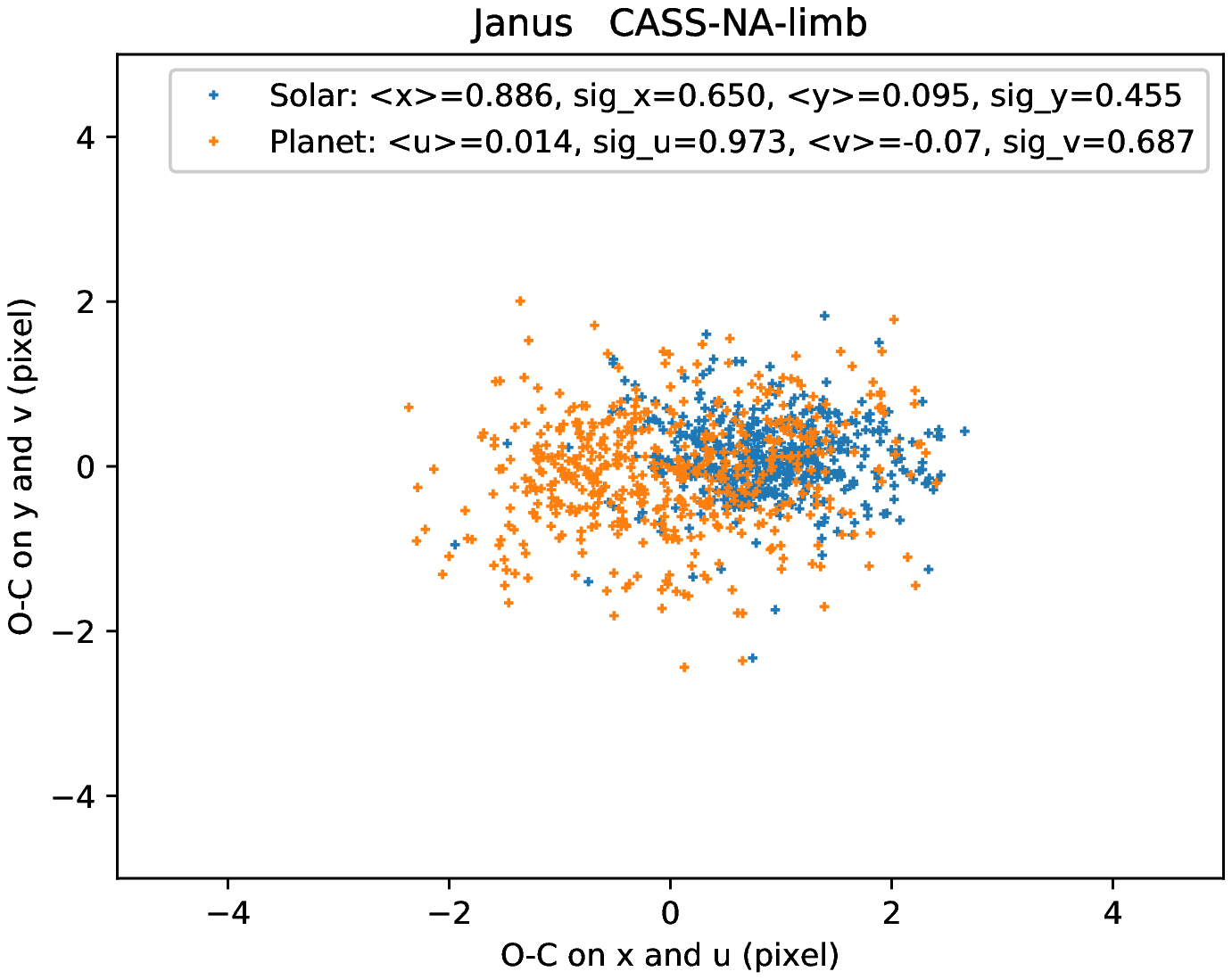}\\
\hspace*{-0.35cm}\includegraphics[width=7.5cm,height=6.cm,angle=0]{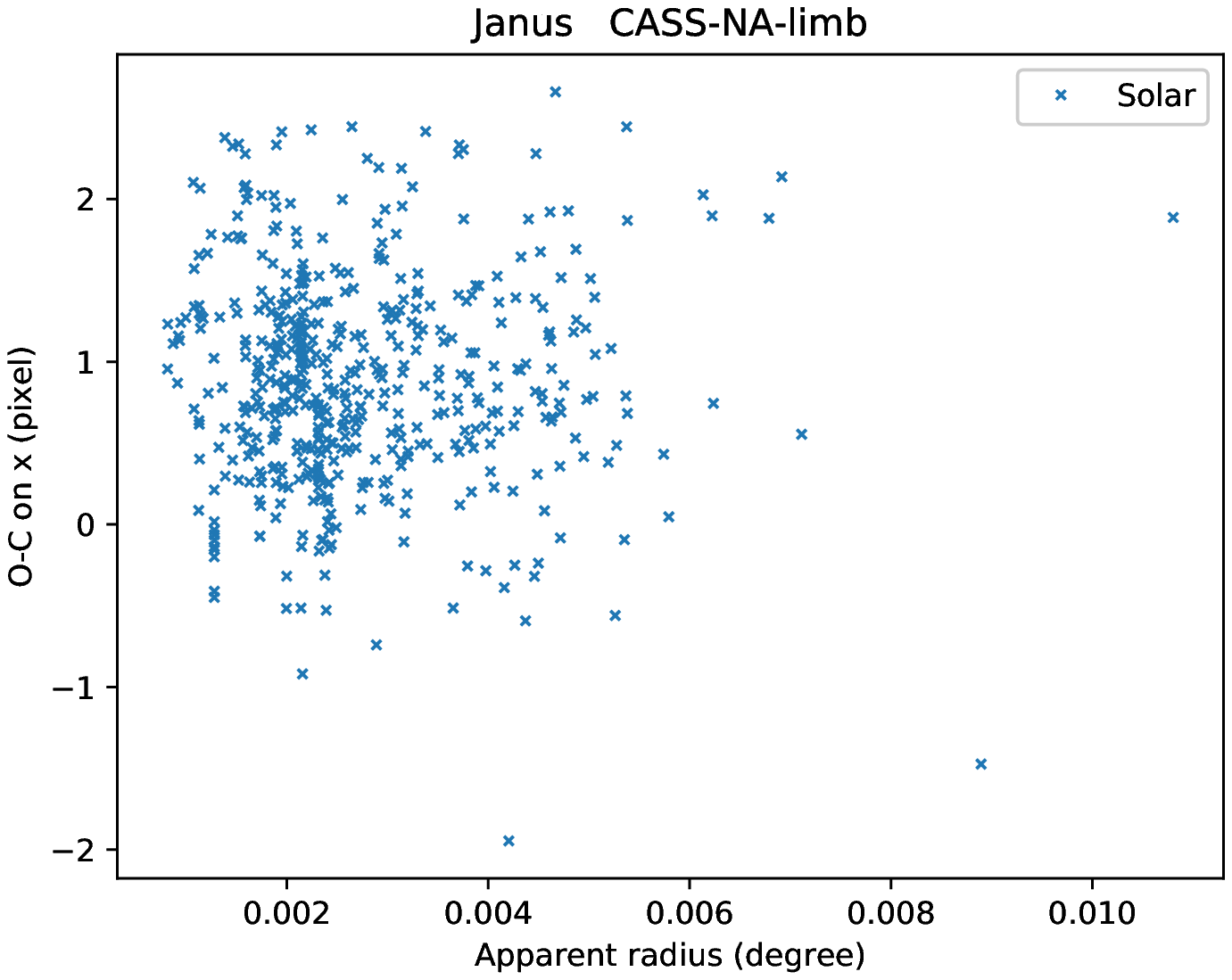} & \includegraphics[width=7.5cm,height=6.cm,angle=0]{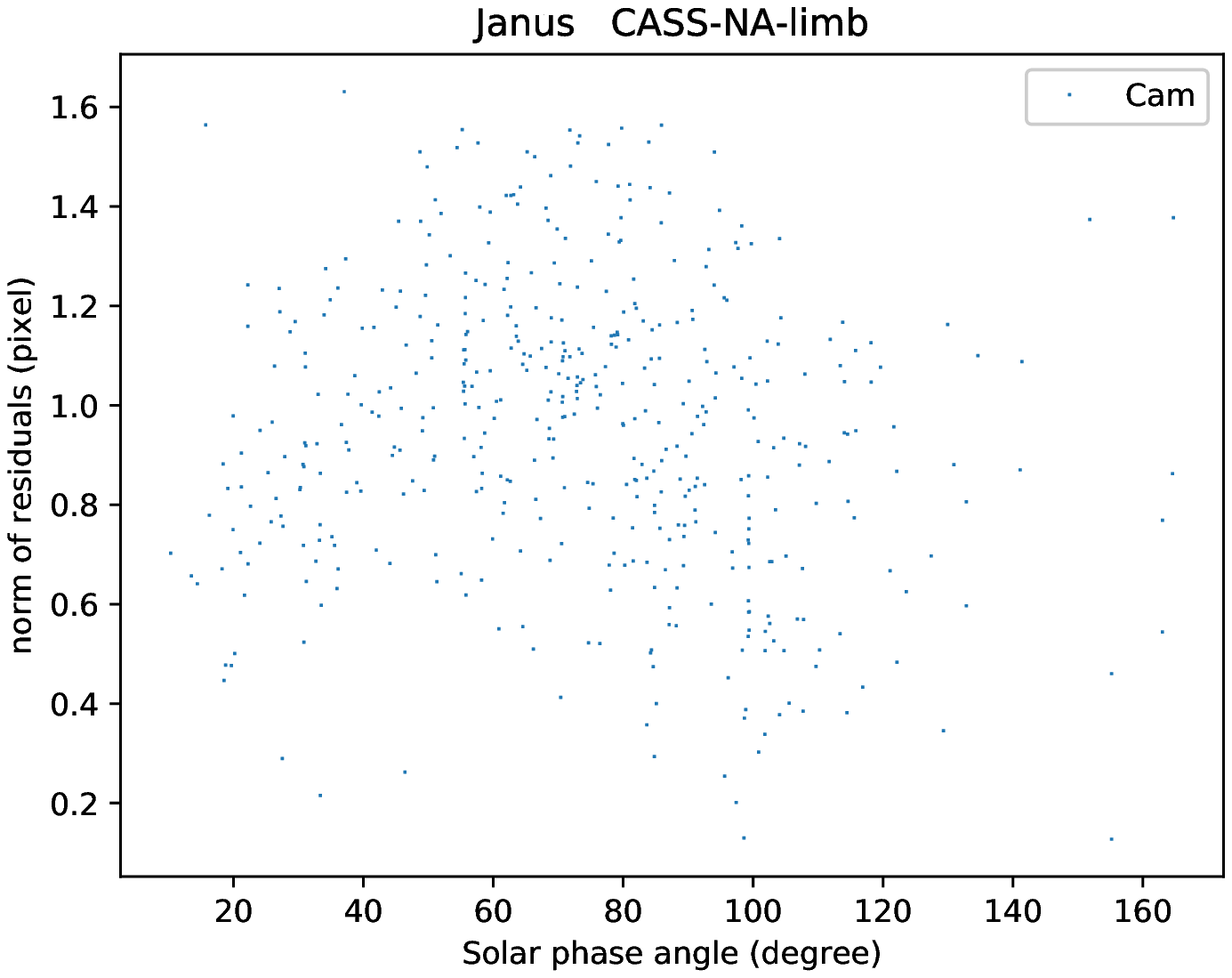}
\end{tabular} 
\caption{Solar bias appearing on Janus' Cassini-ISS data. On the top are given the astrometric residuals in native sample and line coordinates (left, orange color), $x, y$ coordinates where $x$ pointes toward the Sun (left and right, blue color) and $u, v$ coordinates where $u$ points toward Saturn (right, orange color). Residuals along the Solar direction are shown as function of Janus' apparent radius on the images (bottom, left). Norm of residuals as function of Solar phase angle is given at the bottom on the right.}\label{fig:solarbias}
\end{center} 
\end{figure*}

While \citet{Coop14} worked mainly with images where the satellite's center of figure was estimated by a limb fitting method, here both limb fitting and centroiding were used in the data reduction.
A centroiding method was used for Atlas, since it has a small spatial extent on most images, while Prometheus, Pandora, Janus and Epimetheus were mainly treated with limb fitting. 
Having removed the observations performed in the arrival phase of the mission which are essentially biased by the uncertainty in the Cassini spacecraft position, all NAC data were split into two different sets: limb fitting vs centroiding.
Astrometric residuals were then computed in native sample/line $s, l$ coordinates, but also in $x, y$ coordinates where $x$ points toward the Sun and $u, v$ coordinates where $u$ points toward Saturn. We show in Figure \ref{fig:solarbias} an example of such plots in the case of Janus. While Janus provided the largest biases among the five moons we considered, it is noteworthy that all satellites showed exactly the same trend. 
A first conclusion was that the bias is essentially along the Solar direction. Such a conclusion is in full agreement with what was observed for Mimas by \citet{Taje13} and \citet{Coop14} for the main moons.
Since the moons are in spin orbit resonance, and because of the geometric configuration during images capture, the solar bias could in reality be a center of figure / center of mass shift. 
We compared the residuals in Solar $x, y$ coordinates and in planet's $u, v$ coordinates. Contrary to the $x$ coordinate, we did not find any clear signal along the $u$ direction, confirming the solar phase origin of the bias. 
\newline

In a second step, we tried assessing a possible dependency of the Solar bias as a function of the satellite distance, phase angle and mean anomaly. In all three cases, no clear dependency was observed (see Figure \ref{fig:solarbias}).
For the distance dependency, this was a bit surprising since the further the moons, the smaller the bias might be. Nevertheless, it is important to recall that limb fitting techniques require two independent steps: i) edge detection; ii) ellipse fitting. Hence, when the moons appear as small objects on the images, their shapes are poorly constrained since only a small number of points can be detected by the edge detection technique. Such a low number of surface points counterbalances the expected smaller solar bias effect when introducing the ellipse fitting method. This may explain why no obvious dependency of the solar bias on the satellite's distance was found. 
\newline

Since no dependency was found for the solar bias, we found it to be efficient simply to remove a constant bias on the $x$ coordinate for each astrometric data point. Interestingly, a majority of images were taken with the Sun being along the line direction. This explains why no similar large bias arose in the sample direction. Hence, \citet{Lain17} removed a constant bias on the line coordinate, only. Nevertheless, our method is better since the solar angle may be different depending on the images. We determined that the bias to be removed along the $x$ coordinate was 0.89, 0.81, 0.70, 0.72 and 0.49 pixel for Janus, Epimetheus, Pandora, Prometheus and Atlas, respectively.
\newline

Contrary to the other moons, Atlas was essentially treated using a centroiding technique. Similar tests were performed with the same conclusions, except that a bias of 0.3 pixel was found with opposite sign (0.3 has to be added to the $x$ quantity). We should emphasize that a centroiding technique with the Caviar software automatically computed a phase correction. In the case of Atlas, it is likely that such a correction depending on shape and scattering law was not properly modeled.

\section{Gaussian behavior}\label{app:2}

Many estimations of physical quantities in astrophysics generally assume a Gaussian law for modeling the observation noise. Results are then given as standard deviation times some integer (generally 1 or 3).
Assuming the noise to be exactly Gaussian, it implies that a fitted physical value has about 66\% to be within $1 \sigma$ and more than 99\% to be within $3 \sigma$. It is noteworthy to recall that these numbers are correct {\it only} if one assumes a perfect physical modeling and weight procedure, too. 
\newline

\begin{figure*} 
\begin{center} 
\begin{tabular}{ll} 
\hspace*{-0.35cm}\includegraphics[width=7.5cm,height=7.5cm,angle=0]{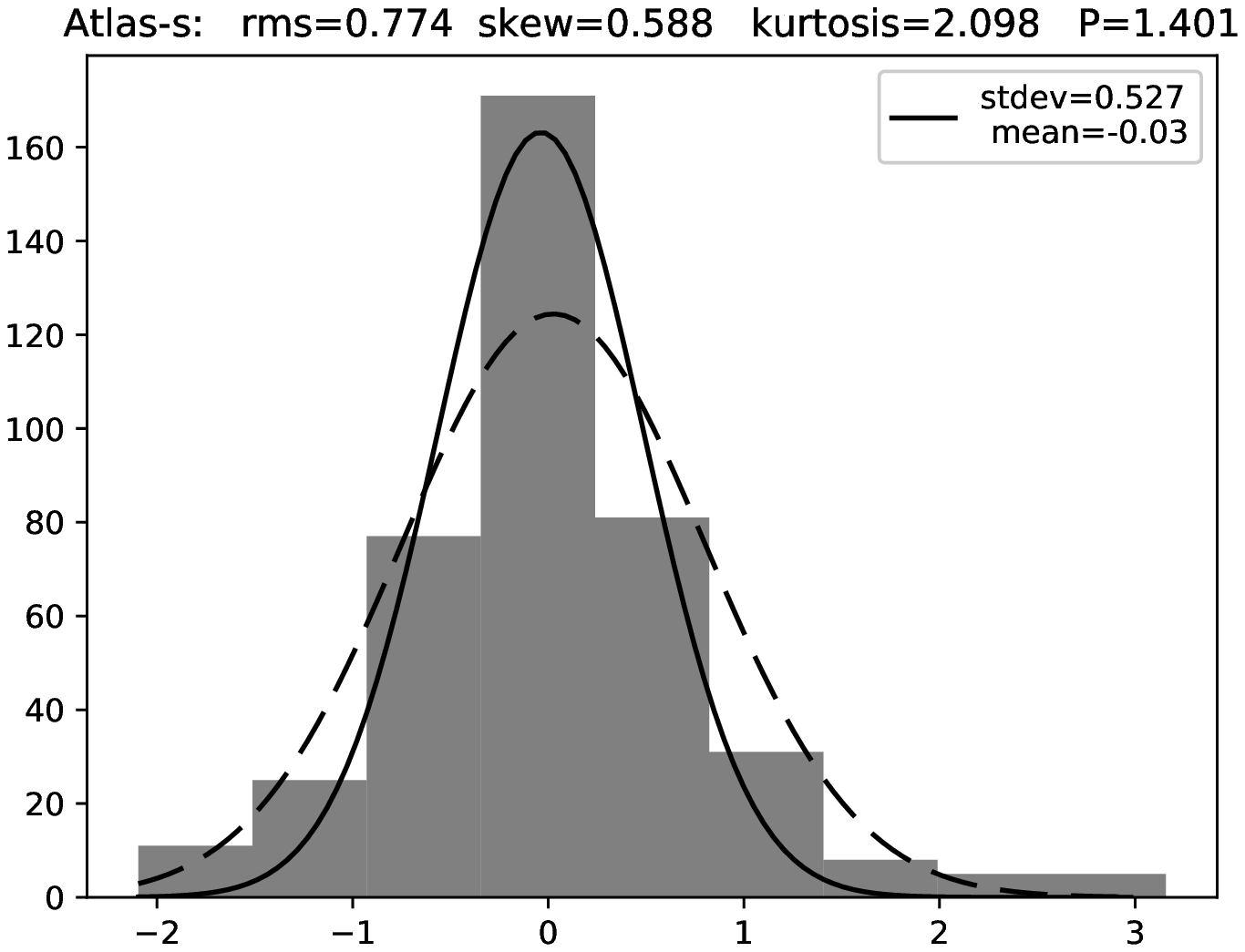} & \includegraphics[width=7.5cm,height=7.5cm,angle=0]{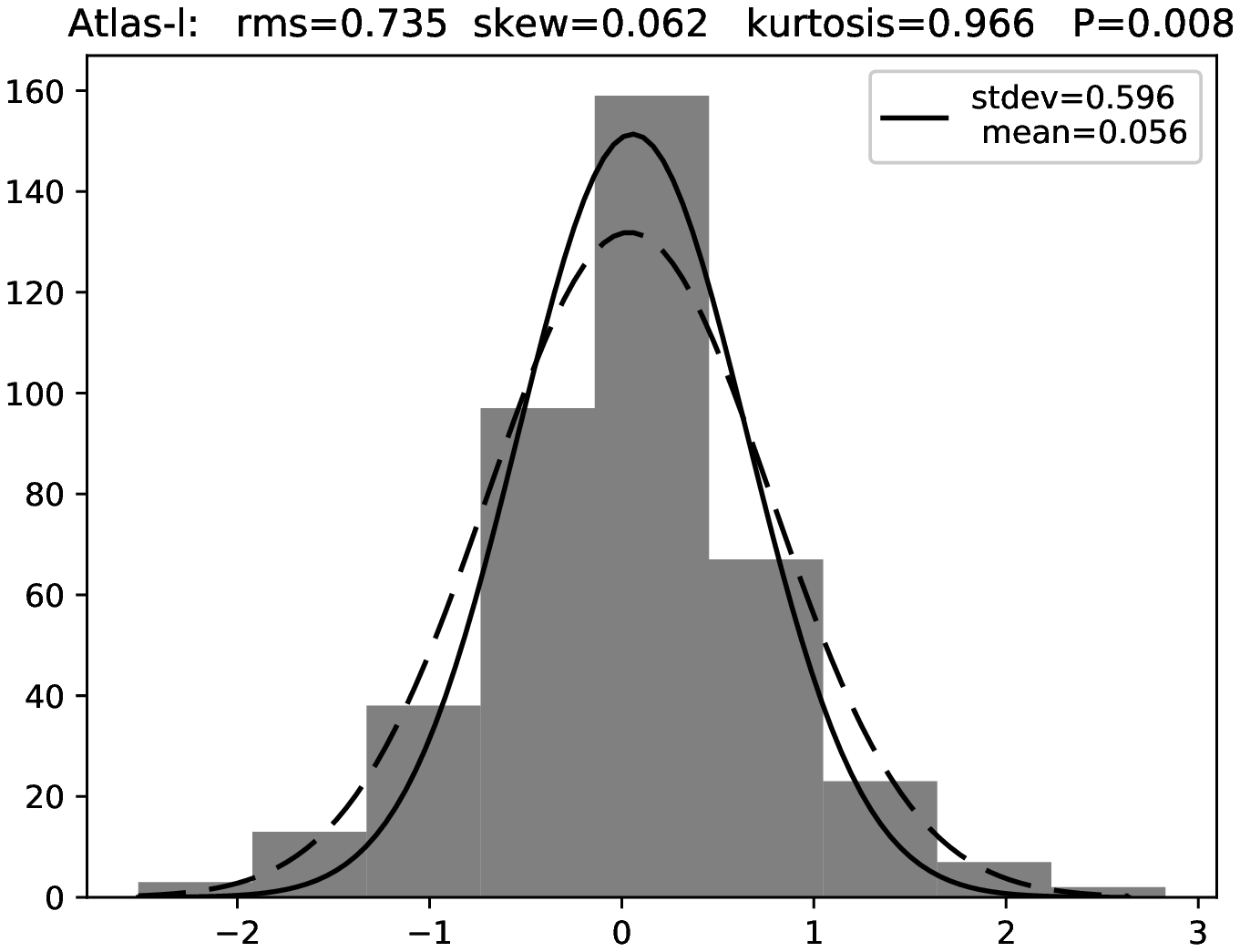}
\end{tabular} 
\caption{Histogram of astrometric residuals for Atlas images treated by centroiding method finding after the Cassini arrival phase.}\label{fig:histCASS-NA-cntr}
\end{center} 
\end{figure*}

\begin{figure*} 
\begin{center} 
\begin{tabular}{ll} 
\hspace*{-0.35cm}\includegraphics[width=7.5cm,height=7.5cm,angle=0]{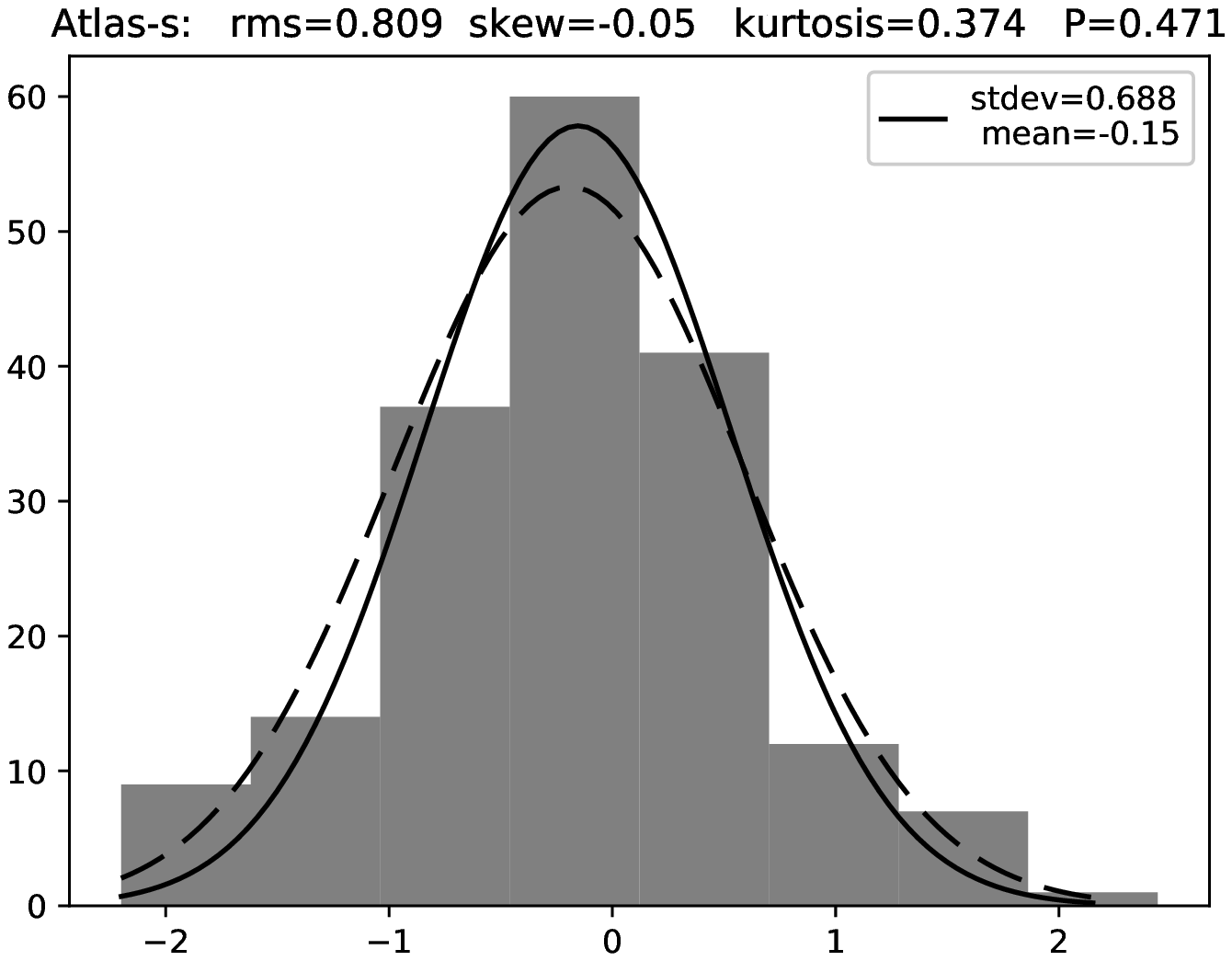} & \includegraphics[width=7.5cm,height=7.5cm,angle=0]{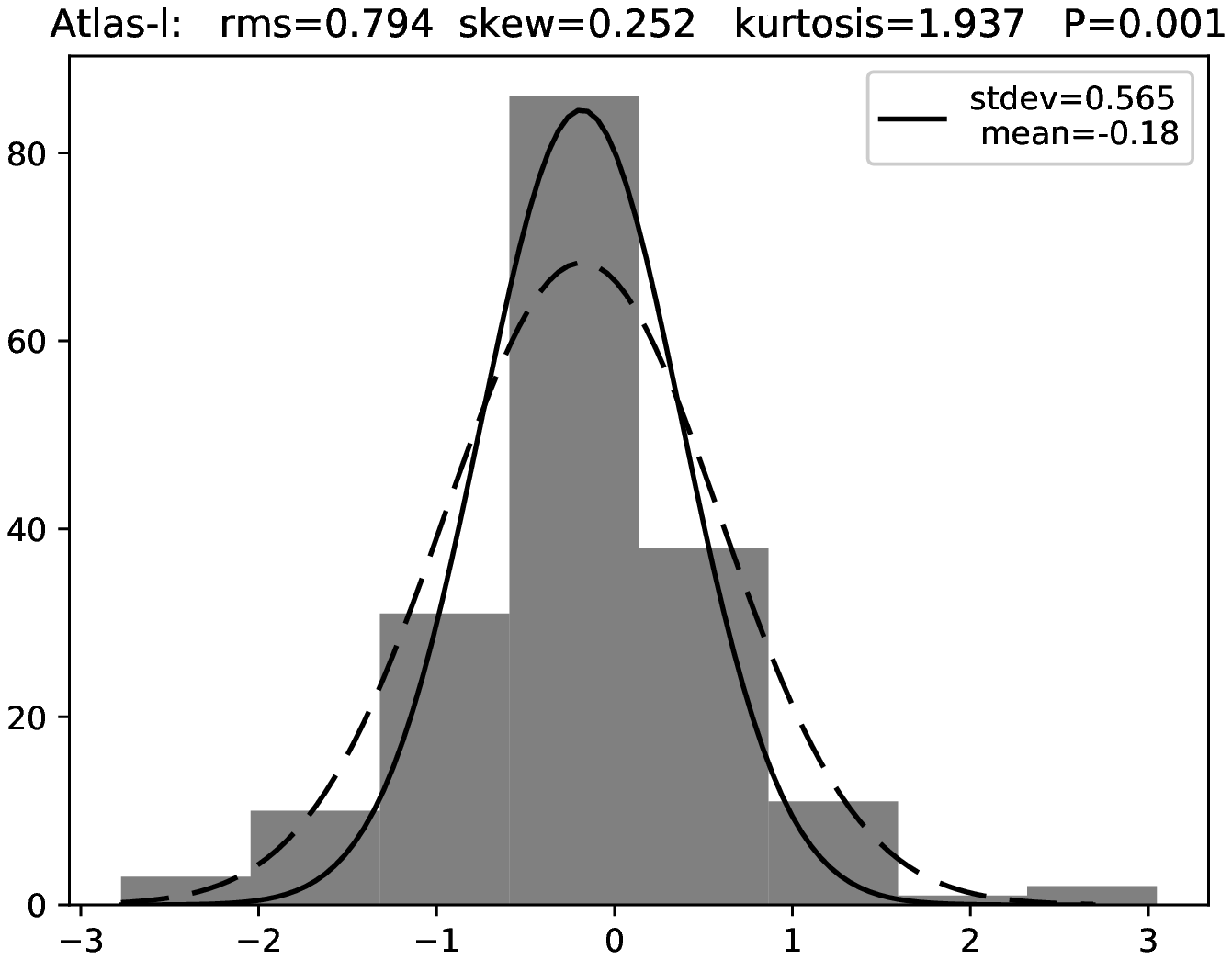}\\
\hspace*{-0.35cm}\includegraphics[width=7.5cm,height=7.5cm,angle=0]{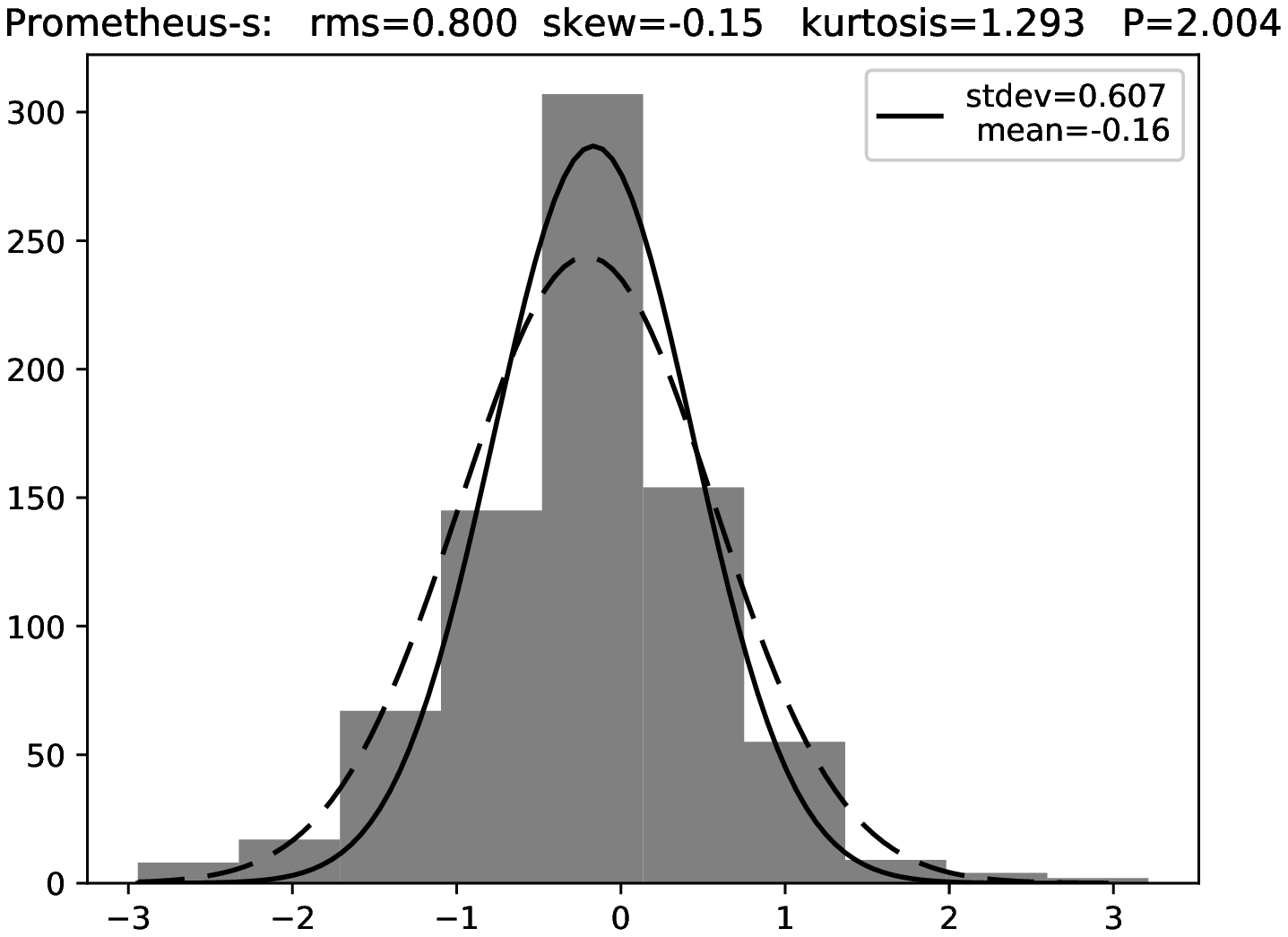} & \includegraphics[width=7.5cm,height=7.5cm,angle=0]{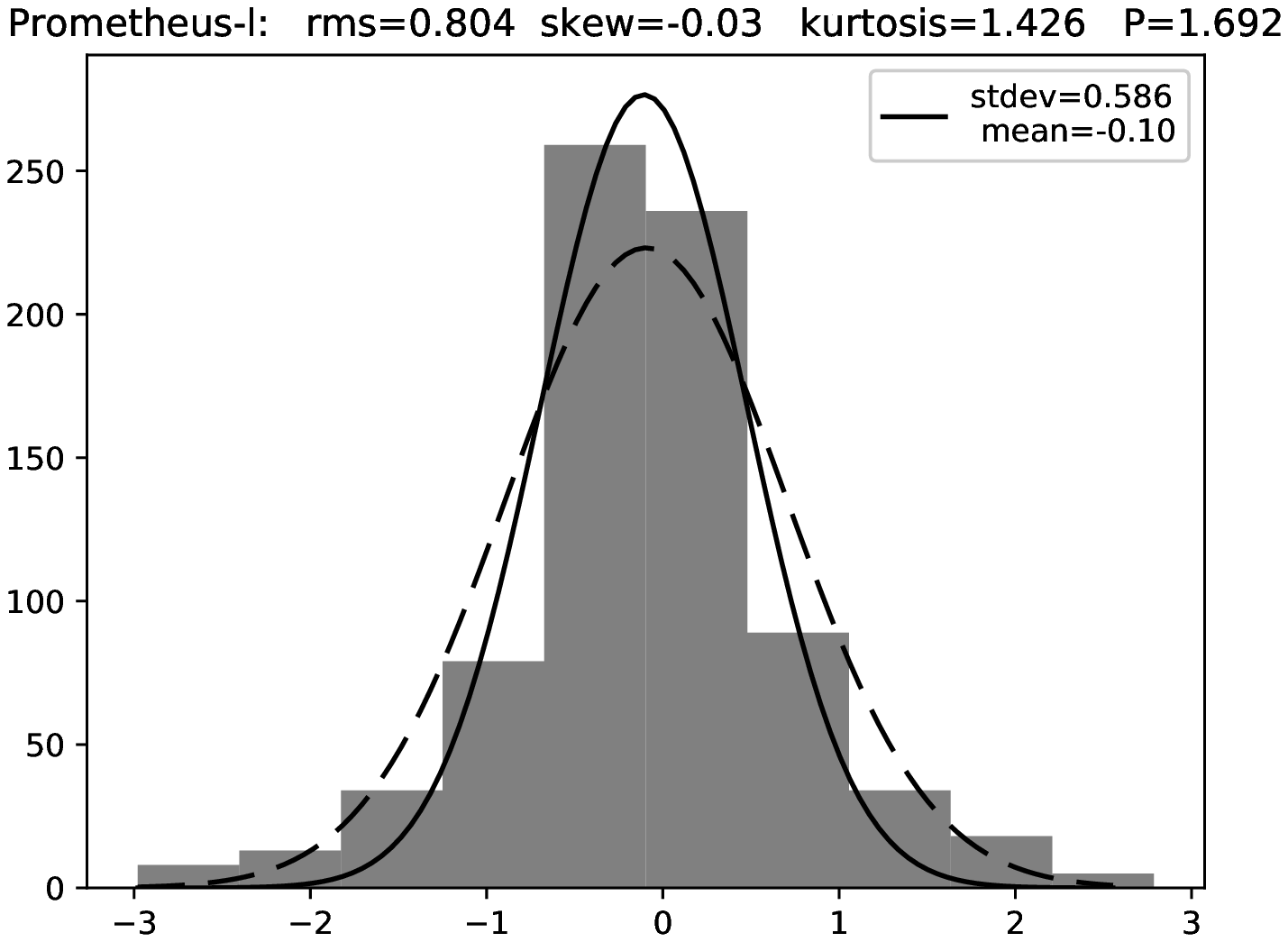}\\
\hspace*{-0.35cm}\includegraphics[width=7.5cm,height=7.5cm,angle=0]{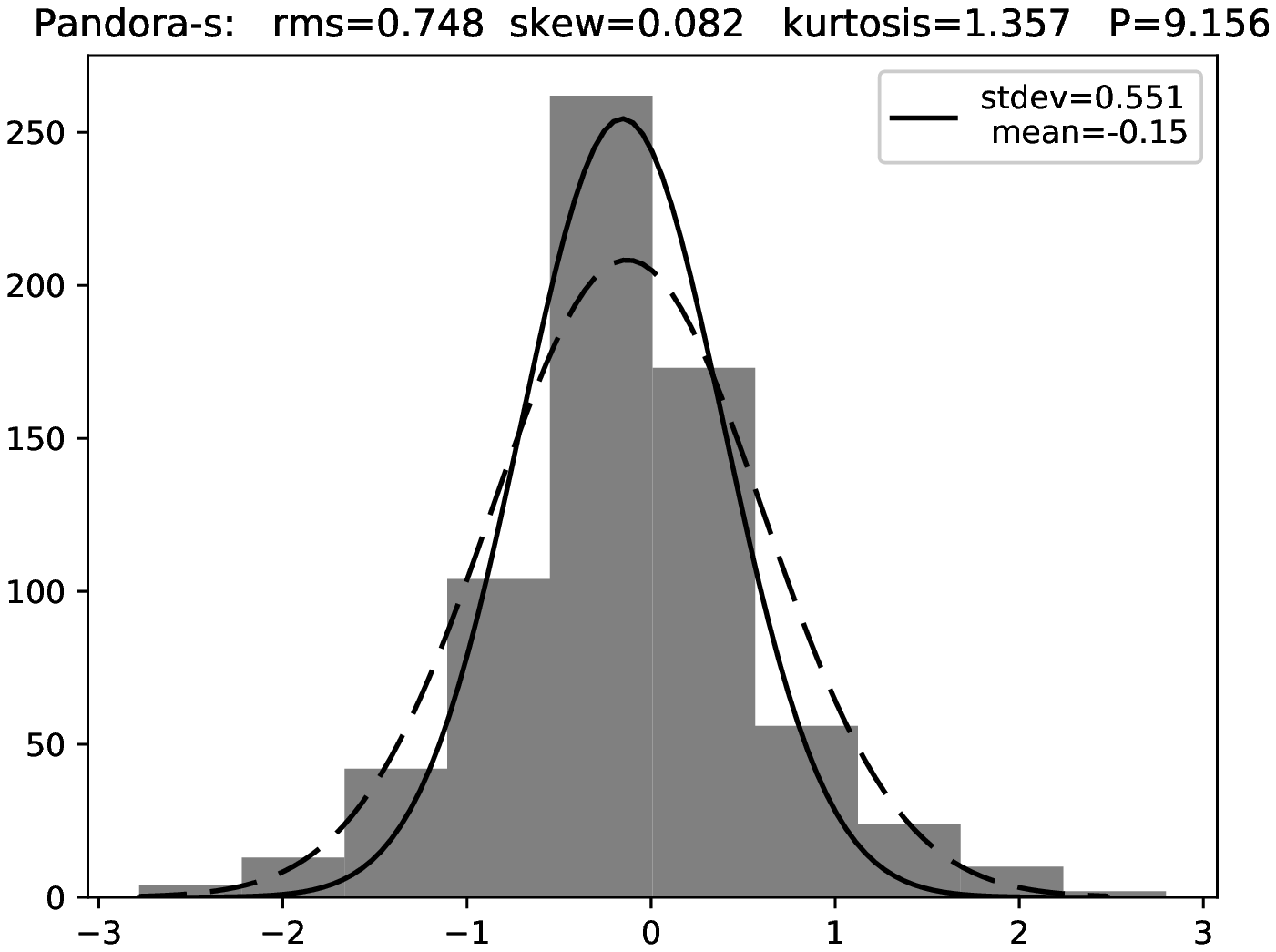} & \includegraphics[width=7.5cm,height=7.5cm,angle=0]{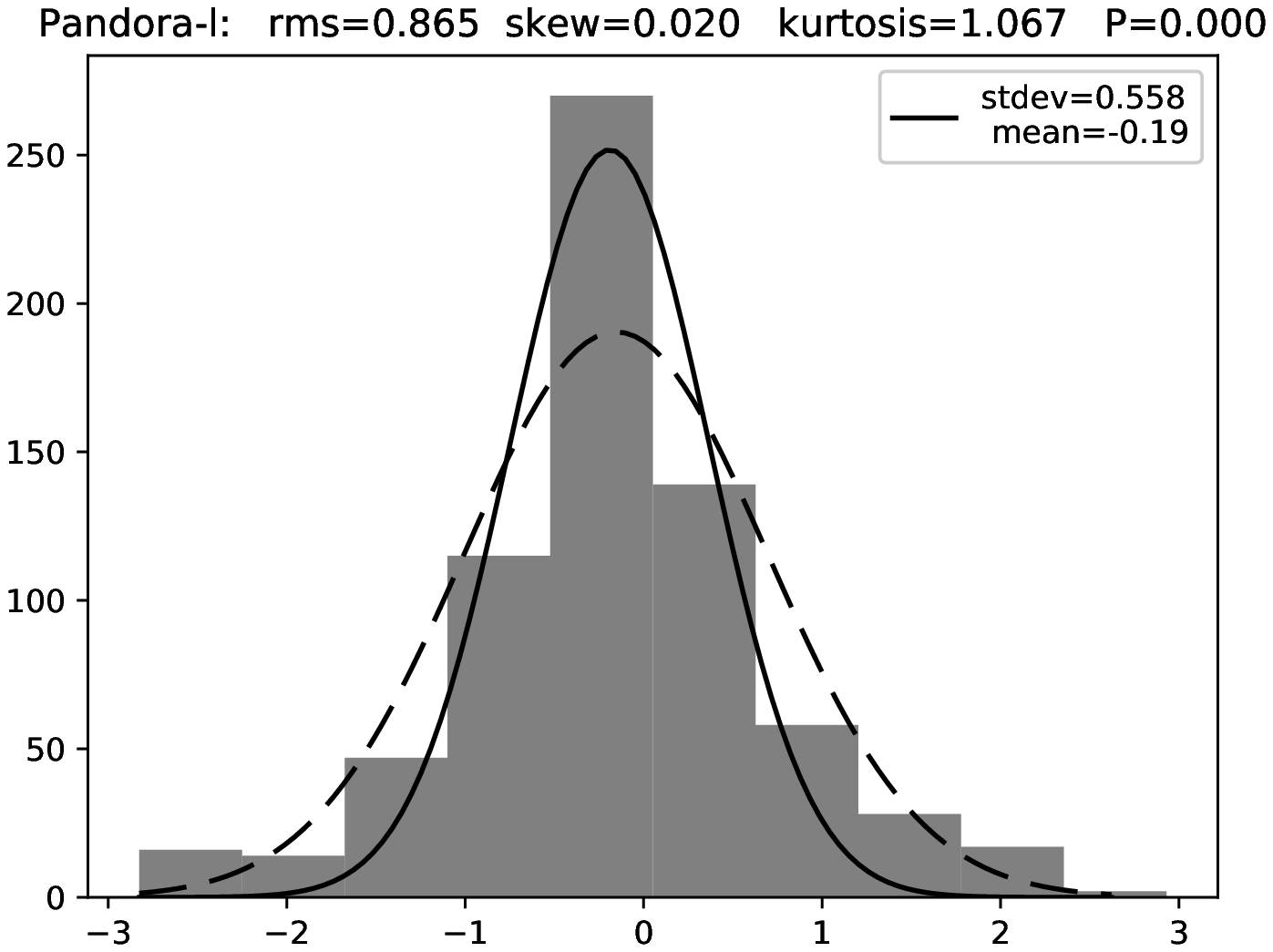}
\end{tabular} 
\caption{Histogram of astrometric residuals for Atlas, Prometheus and Pandora images treated by limb fitting method.}\label{fig:histCASS-NA-limb}
\end{center} 
\end{figure*}

\begin{figure*} 
\begin{center} 
\begin{tabular}{ll} 
\hspace*{-0.35cm}\includegraphics[width=7.5cm,height=7.5cm,angle=0]{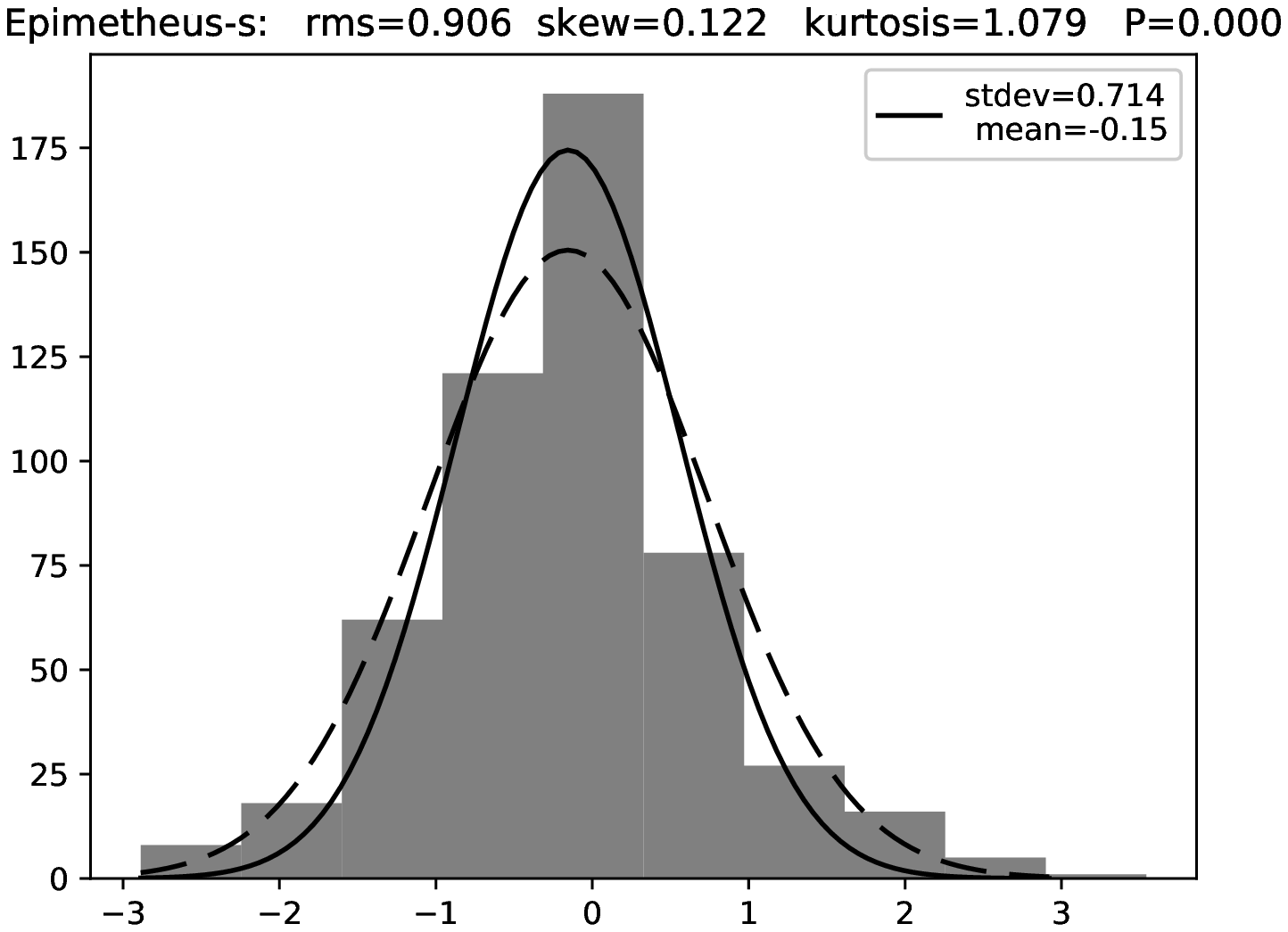} & \includegraphics[width=7.5cm,height=7.5cm,angle=0]{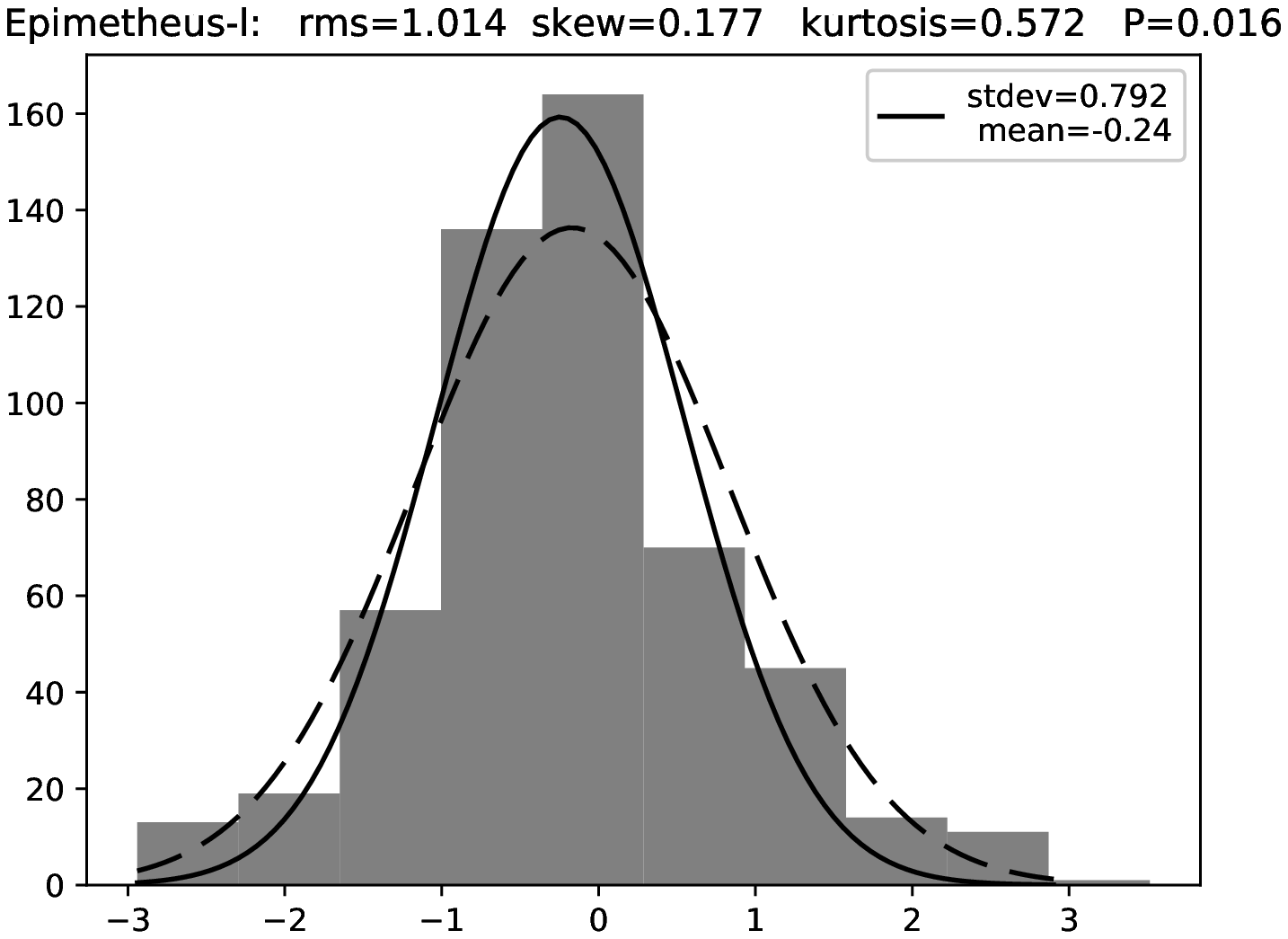}\\
\hspace*{-0.35cm}\includegraphics[width=7.5cm,height=7.5cm,angle=0]{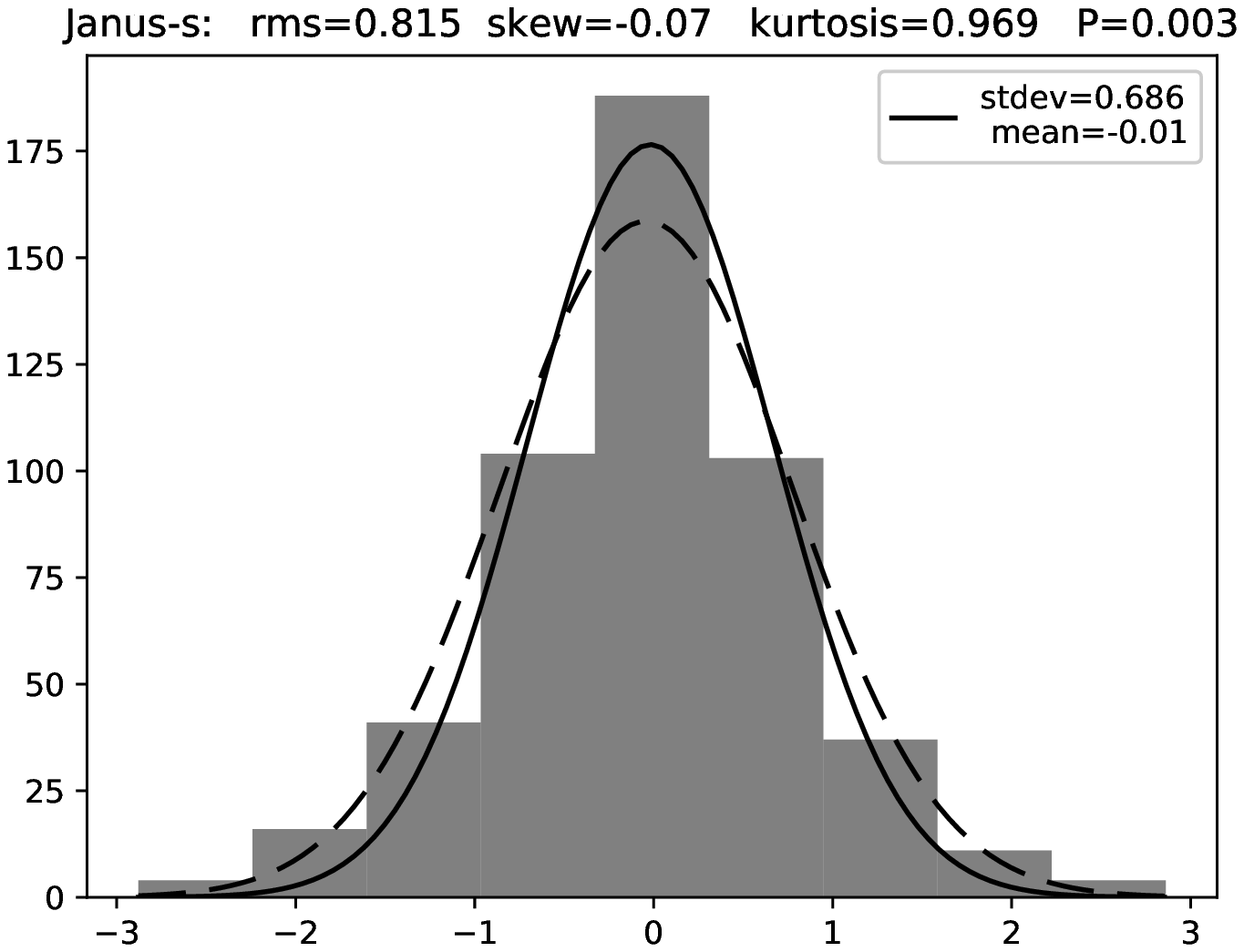} & \includegraphics[width=7.5cm,height=7.5cm,angle=0]{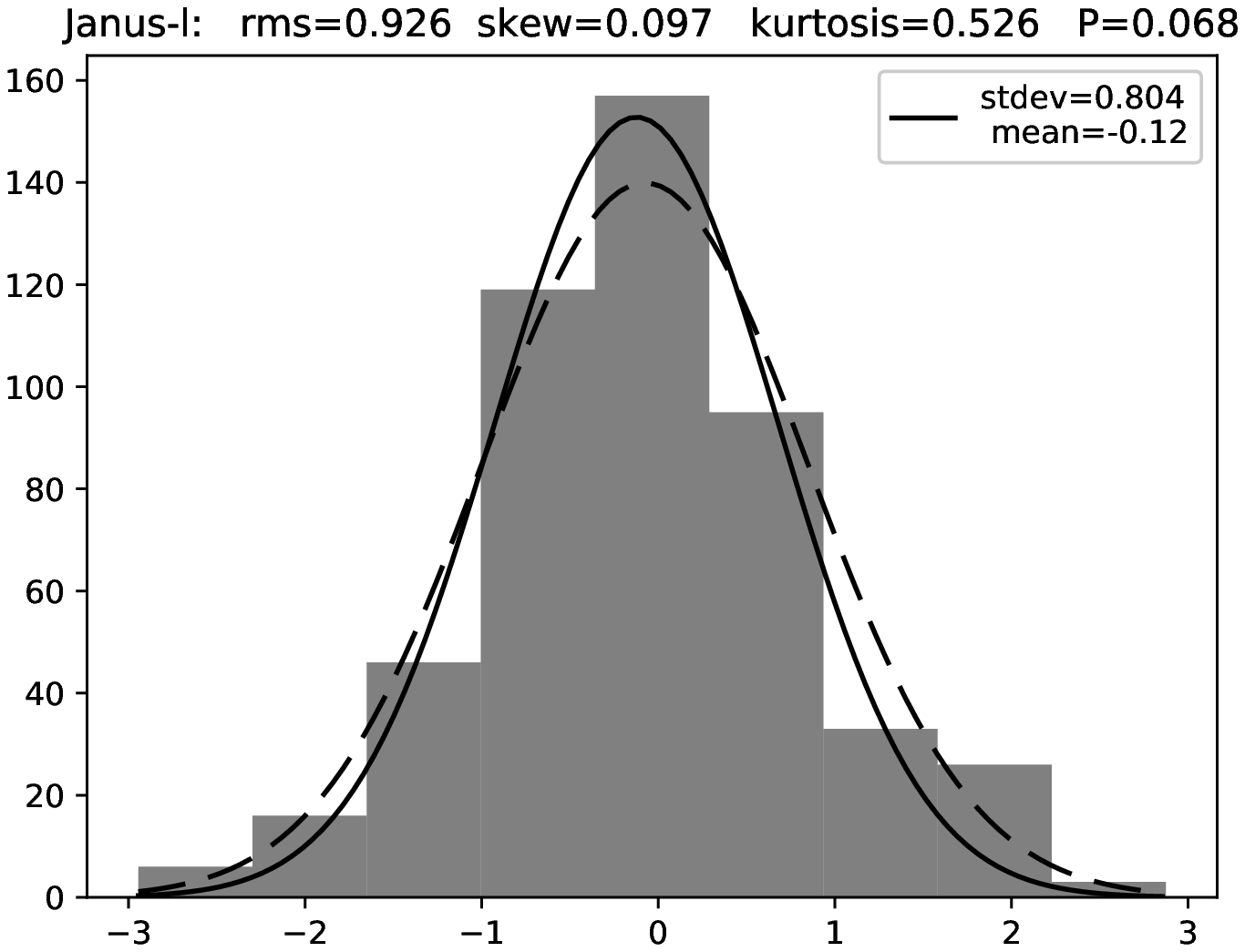}
\end{tabular} 
\caption{Histogram of astrometric residuals for Epimetheus and Janus images treated by limb fitting method.}\label{fig:histCASS-NA-limb2}
\end{center} 
\end{figure*}

Here we tried to assess if the noise on the ISS-NAC astrometric data is close to Gaussian. Similarly to what is explained in the main text, we split the data into subsets considering the camera, the center of figure algorithm and separating the observations during arrival phase. Then all subsets were compared to a Gaussian distribution. Since in our fitting procedure, every data point was considered with a specific weight determined as a combination of measurement uncertainty and rescale parameters (see main text for details), all data were divided here by such a weight before being treated in our Gaussian estimation. This was a requirement since each data get its own uncertainty from Caviar, obtained from specific observation conditions.
As a consequence, a perfect Gaussian behavior should provide a standard deviation of 1. In Figures \ref{fig:histCASS-NA-cntr} to \ref{fig:histCASS-NA-limb2} we provide histograms of our astrometric residuals after rescaling the data by their respective weight. It appears that the standard deviations are significantly lower than unity. Since, our plots were obtained after removal of the Solar bias only, in particular without any refitting and re-estimation of the rescaling parameters, the significant departure from unity shows the gain obtained from treating properly the Solar bias. In addition the skewness and kurtosis, two indicators of Gaussian shape, are given with the D'Agostino and Pearson estimator. For both quantities, a Gaussian behavior should be associated with a number close to zero. Last, we provide on all Figures two Gaussian curves. The dashed one was determined using mean and standard deviation directly computed from the data set, while the full one was obtained after fitting a Gaussian to the histogram. Our plots support the hypothesis that the noise in our astrometric data set, is close to a Gaussian distribution. Moreover, we clearly see that the controiding and limb fitting methods do not differ from each other in terms of the statistical distribution of residuals.
\newline

\section{How the libration affects the precession of the pericenter}\label{app:derivation}

\par This section is widely inspired from \citep{Jaco10}. The potential for the acceleration of Saturn due to the quadrupole potential of the satellite is

\begin{equation}
	\label{eq:potentiel}
	\mathcal{U}' = \frac{Gm}{2}\frac{R^2}{r^3}\left(J_2+6C_{22}\cos 2\theta\right),
\end{equation}
where $m$ is the mass of the satellite, $R$ its mean equatorial radius, $J_2$ and $C_{22}$ its un-normalized gravity coefficients, and $r$ the distance Saturn-satellite. $\theta = -\psi$ is the angle between the direction from the satellite to Saturn, and the axis of minimum principal moment of inertia. It reads

\begin{equation}
	\label{eq:theta}
	\theta = f-M-\mathcal{A}\sin(M).
\end{equation}

Following the action-reaction law, the potential acting on the satellite is $\mathcal{U}=-M_{\saturn}/m\mathcal{U}'$, where $M_{\saturn}$ is the mass of Saturn, and from the classical formulae

\begin{eqnarray}
	\mathcal{G}M_{\saturn} & = & n^2a^3, \label{eq:kepler} \\
	r & = & a\left(1+\frac{e^2}{2}-2e\sum_{s=1}^{\infty}\frac{1}{s^2}\frac{d}{de}J_s(se)\cos sM\right), \label{eq:ra} \\
	f-M & = & 2e\sin M+\frac{5}{4}e^2\sin 2M+O(e^3), \label{eq:fM} \\
	\frac{d\varpi}{dt} & = & \frac{\sqrt{1-e^2}}{na^2e}\frac{\partial\mathcal{U}}{\partial e}, \label{eq:laplacelagrange}
\end{eqnarray}
we get

\begin{equation}
	\label{eq:dvarpidt}
	\frac{d\varpi}{dt} = \frac{3}{2}n\left(\frac{R}{a}\right)^2\left[J_2-2C_{22}\left(5-\frac{4\mathcal{A}}{e}\right)+\frac{J_2+6C_{22}}{e}\cos M\right]+O(e),
\end{equation}
that is straightforwardly integrated as

\begin{equation}
	\label{eq:deltavarpi}
	\Delta\varpi = \frac{3}{2}\left(\frac{R}{a}\right)^2\left[J_2-2C_{22}\left(5-\frac{4\mathcal{A}}{e}\right)\right]nt+\frac{3}{2}\left(\frac{R}{a}\right)^2\frac{J_2+6C_{22}}{e}\sin M.
\end{equation}


\section*{Acknowledgements}

VL acknowledges W. Folkner for suggesting some of the testing method presented in this paper and R. Jacobson for former exchanges about the Phobos physical libration. The authors are extremely grateful to the two referees whose comments and suggesrtions much improved this paper. We thank S. Le Maistre for his reading and suggestions, also. VL's research was supported by an appointment to the NASA Postdoctoral Program at the NASA Jet Propulsion Laboratory, California Institute of Technology, administered by Universities Space Research Association under contract with NASA. The authors are indebted to all participants of the Encelade WG. This work has been supported by the European Community's Seventh Framework Program (FP7/2007-2013) under grant agreement 263466 for the FP7-ESPaCE project and the International Space Science Institute (ISSI). B.N. acknowledges support of the contract Prodex CR90253 from the Belgian Science Policy Office (BELSPO).
N.C. and C.M. were supported by the UK Science and Technology Facilities Council (Grant No. ST/M001202/1) and are grateful to them for financial assistance. C.M. is also grateful to the Leverhulme Trust for the award of a Research Fellowship. N.C. thanks the Scientific Council of the Paris Observatory for funding.

 \end{document}